\documentclass[fleqn, usenatbib]{mnras}
\usepackage{graphicx}	
\usepackage{amsmath}	
\usepackage{amssymb}	
\usepackage{multicol}
\usepackage[T1]{fontenc}
\usepackage{ae,aecompl}

\usepackage{txfonts}

\title[Modelling Local Group Supernova Remnants]{Supernova Remnants in the Local Group I: A model for the radio luminosity function and visibility times of supernova remnants}

\author[S. K. Sarbadhicary, , C. Badenes, L. Chomiuk, D. Caprioli]{Sumit K. Sarbadhicary$^{1}$\thanks{sks67@pitt.edu}, Carles Badenes$^{1}$, Laura Chomiuk$^{2}$, Damiano Caprioli$^{3}$,
\newauthor and Daniel Huizenga$^{2}$\\
\\
$^{1}$\small Department of Physics and Astronomy and Pittsburgh Particle Physics, Astrophysics and Cosmology Center (PITT PACC), University of Pittsburgh, 3941\\
O`Hara St, Pittsburgh, PA 15260, USA\\
$^{2}$\small Department of Physics and Astronomy, Michigan State University, East Lansing, MI 48824, USA\\
$^{3}$\small Department of Astrophysical Sciences, Princeton University, Ivy Lane, Princeton, NJ 08540, USA}

\begin{document}
\maketitle
\begin{abstract}
Supernova remnants (SNRs) in Local Group galaxies offer unique insights into the origin of different types of supernovae. In order to take full advantage of these insights, one must understand the intrinsic and environmental diversity of SNRs in the context of their host galaxies. We introduce a semi-analytic model that reproduces the statistical properties of a radio continuum-selected SNR population, taking into account the detection limits of radio surveys, the range of SN kinetic energies, the measured ISM and stellar mass distribution in the host galaxy from multi-wavelength images and the current understanding of electron acceleration and magnetic field amplification in SNR shocks from first-principle kinetic simulations. Applying our model to the SNR population in M33, we reproduce the SNR radio luminosity function  with a median SN rate of $\sim 3.1 \times 10^{-3}$ per year and an electron acceleration efficiency, $\epsilon_{\rm{e}} \sim 4.2 \times 10^{-3}$. We predict that the radio visibility times of $\sim 70\%$ of M33 SNRs will be determined by their Sedov-Taylor lifetimes, and correlated with the measured ISM column density, $N_H$ ($t_{\rm{vis}} \propto N_H^{-a}$, with $a \sim 0.33$) while the remaining will have visibility times determined by the detection limit of the radio survey. These observational constraints on the visibility time of SNRs will allow us to use SNR catalogs as `SN surveys' to calculate SN rates and delay time distributions in the Local Group.
\end{abstract}


\section{Introduction}
Supernova remnants (SNRs) contain clues to the nature of progenitors of Type Ia and core-collapse (CC) SNe. For Type Ia SNe, the two leading progenitor models are the single degenerate (white dwarf with a non-degenerate star) and double degenerate (a pair of white dwarfs) scenarios, neither of which meets all current theoretical and observational constraints \protect\citep[See][for reviews]{Wang2012, Maoz2014b}. For CC SNe, the progenitors are better constrained as explosions of stars $\gtrsim 7-8 \rm{M_{\odot}}$ \protect\citep{Smartt2009, Jennings2012, Jennings2014}. The observed deficit of red supergiant progenitors between $18-30\ \rm{M_{\odot}}$ \protect\citep{Kochanek2008, Smartt2009a} however, could suggest that CC progenitors in this mass range may directly produce black holes without a visible supernova \protect\citep[see][for a review]{Smartt2015}, and most theoretical models still do not include the possibly large effects of binary interactions \protect\citep[][ Zapartas et al, in prep]{Sana2012}. Signatures of different progenitor pathways can be found in SNRs in the Local Group, which can be studied in great detail thanks to advances in imaging and spectroscopy. For example, the single and double degenerate channels of Type Ia have been tested using the morphology and X-ray spectra of known Type Ia SNRs \protect\citep{Badenes2007, Badenes2010a, Vink2011a, Yamaguchi2015} and by looking for the presence of surviving companion stars near the SNR centers \protect\citep{Canal2001, Edwards2012, Schaefer2012, Pagnotta2015}. However, these techniques are limited to the handful of young, ejecta-dominated SNRs, whereas the bulk of SNRs in Local Group galaxies are older objects in the Sedov stage.  

\protect\cite{Badenes2010} and \protect\cite{Maoz2010} tackled the SN progenitor problem by pioneering the use of SNR populations as `effective' SN surveys to calculate the SN delay-time distribution (DTD). The DTD is the SN rate that would be observed following a hypothetical brief burst of star formation \protect\citep{Maoz2013a}, and can be measured from a SN survey and a set of star formation histories. The DTD serves as a powerful observational constraint on SN progenitor models because it effectively encodes the evolutionary timescales of different progenitor channels. The Type Ia SN DTD has been often measured from extragalactic SN surveys \protect\citep{Totani2008, Maoz2010a, Maoz2011, Graur2011, Maoz2012, Graur2013, Graur2014}, but the integrated spectra of these galaxies yield luminosity-weighted estimates of stellar ages and masses, which can introduce systematic errors in the derived star-formation histories \protect\citep{Wuyts2011, Conroy2013} that can bias the resulting DTD. Local Group galaxies, such as the Magellanic Clouds used by \protect\cite{Maoz2010}, provide more accurate star formation histories from observations of \textit{resolved} stellar populations \protect\citep{Harris2004, Harris2009, Lewis2015}. In these galaxies, an effective SN survey can be conducted using SNRs instead of SNe (which only explode a few times per century in the Local Group). 

However, measuring the SN rate from SNRs requires an estimate of their \textit{visibility time}, the duration for which SNRs in the survey would remain detectable, which is related to the environment-sensitive evolution of SNRs. \protect\cite{Badenes2010} approximated the visibility times of the Magellanic SNR population as the cooling timescale of the SNR plasma transitioning from the Sedov to the radiative stage \protect\citep{Blondin1998, Bandiera2010}. This approximation does not take into account the details of the synchrotron radio light curves of SNRs produced by particle acceleration at the shock-ISM interface. \protect\cite{Badenes2010} also assumed that the heterogenous sample of Magellanic SNRs was complete, whereas radio SNR surveys are known to be sensitivity-limited, thereby possibly missing many faint SNRs \protect\citep{Gordon1999, Chomiuk2009a}. 

In this study, we model the radio visibility times of a catalog of SNRs statistically, by simulating their radio light curves and accounting for completeness of the SNR catalog. Our work is a proof-of-concept, with the goal of extracting physical information relevant to SN studies, such as the SN rate and DTD, from radio continuum-selected SNR surveys and knowledge of the ISM in Local Group galaxies. We model the radio light curves of SNRs using current theories of electron acceleration and magnetic field amplification in SNR shocks \protect\citep[and references therein]{Caprioli2014, Caprioli2014a}. The population of model SNRs is then constrained using multi-wavelength maps of the host galaxy, which trace the ambient ISM and stellar population, and the observed luminosity function (LF) of a radio-continuum selected SNR survey. We focus on the radio properties of SNRs over other wavelengths because: (1) SNRs have been traditionally easier to model in the radio \protect\citep{Chevalier1982a, Berkhuijsen1984, Chevalier1998, Berezhko2004a, Chevalier2006} compared to X-ray or optical wavelengths. X-ray spectra of SNRs is produced by a mix of thermal and non-thermal processes that require complex hydrodynamical modeling \protect\citep{Badenes2010a, Vink2011a}, while the [SII]/H$\alpha$ ratio, commonly used in optical surveys to identify SNRs \protect\citep{Mathewson1973, Gordon1998, Blair2012, Lee2014}, is sensitive to the properties of the interstellar clouds interacting with the shock \protect\citep{McKee1975, Dopita1979, Dopita1984, White1991} and the ionization state of the ambient medium \protect\citep{Morlino2012a}, making it difficult to model. (2) SNR catalogs based on radio synchrotron emission will be largely unaffected by extinction compared to X-ray or optical. Similar statistical studies of Local Group SNR populations have also been done to explain their size distribution of SNRs \protect\citep{Asvarov2014} and constrain the microphysics of interstellar shocks \protect\citep{BarniolDuran2016}.

For this paper, we apply our model to the M33 SNR population since M33 currently has the largest single-survey radio SNR catalog in the Local Group \protect\citep{Gordon1999}, and is free of distance uncertainties, unlike Galactic SNRs \protect\citep{Green2009}. In the future, our model will be extended to M31 and the Magellanic Clouds. In Section \ref{MCSNR}, we describe the generation of model SNR populations using a Monte-Carlo method, with emphasis on our radio light curve model, the strategy for placing SNRs in the ISM, and comparison with the radio SNR catalog in M33. In Section \ref{sec:PARAM}, we show the constraints on the model parameter space from the radio LF of the M33 SNR catalog. Section \ref{vistime} uses these constraints to predict the visibility times of M33 SNRs as a function of the HI column density. Section \ref{sec:BRIGHT} investigates the correlation between the radio luminosity of the brightest SNR and the SN rate, and Section \ref{sec:DISCUSS} discusses the implications of these findings, as well as the effects of changing key assumptions in the model. 

\section{Monte-Carlo model of SNR Populations} \label{MCSNR}
\subsection{Generating SNR populations} \label{genSNRs}
We create synthetic SNR populations by generating multiple SNRs following a Poisson process, where the probability of an SNR exploding in a given year is given by the SN Rate, $R$. We fix the ratio of Type Ia to CC SNe at 1/3 \protect\citep[combining Type Ib/c and Type II SNe into the CC rate, for simplicity]{Li2011}. We wait until the SNR population reaches a `steady state', i.e the rate of SNRs forming and fading roughly balance each other, and then compare the LF of this steady state with observations. Since the steady state is reached within a few $10^4$ years, much smaller than galactic timescales, we keep $R$ and the Ia/CC fraction constant. 

For each SNR, we select a spatial location in M33, an ambient density at that location, and the kinetic energy and ejecta mass that serve as initial conditions for the SNR radio light curve. 

\begin{table*} 
\caption{ Multi-wavelength data for M33 to constrain the SNR evolution model}
\label{multiwavtable}
\centering
\begin{tabular}{ll@{}rrclcl}
\hline
Waveband $^a$& Facility & Angular & Spatial $^b$ & Program & Utility & Reference \\
 & & Resolution (") & Resolution (pc) & & & \\
 \hline
 Radio (21 cm) & VLA + GBT & 20.0 & 77.0 & - & Opacity-corrected HI column density (ISM) & 1 \\ 
 Radio (6, 20 cm) & VLA + WSRT & 7.0 & 29.0 & - & Radio continuum-selected SNR catalog & 2\\
 R (658 nm) & 4 m Mayall/Mosaic & 1.0 & 4.0 & LGGS & Bulk stellar population for placing Type Ia SNRs & 3\\
 FUV (1539 \AA) & GALEX & 4.2 & 17.1& NGS & Star forming regions for placing CC SNRs & 4,5 \\
 IR (24 $\mu$m) & Spitzer/MIPS & 6.0 & 24.4 & - & Star forming regions obscured by dust & 6\\
 \hline
 \multicolumn{7}{l}{$^a$ Median wavelength quoted in parentheses} \\
 \multicolumn{7}{l}{$^b$ Assuming a distance to M33 = 840 kpc \protect\citep{KennicuttJr.2008}}\\
 \multicolumn{7}{l}{\textbf{References}: (1) \protect\cite{Braun2012} (2) \protect\cite{Gordon1999} (3) \protect\cite{Massey2006} (4) \protect\cite{GildePaz2007} (5) \protect\cite{Morrissey2007} (6) \protect\cite{Dale2009} } \\
\end{tabular}
\end{table*}

\subsubsection{Spatial Location} \label{subsec:spat}
\begin{figure*}
\includegraphics[width=\textwidth]{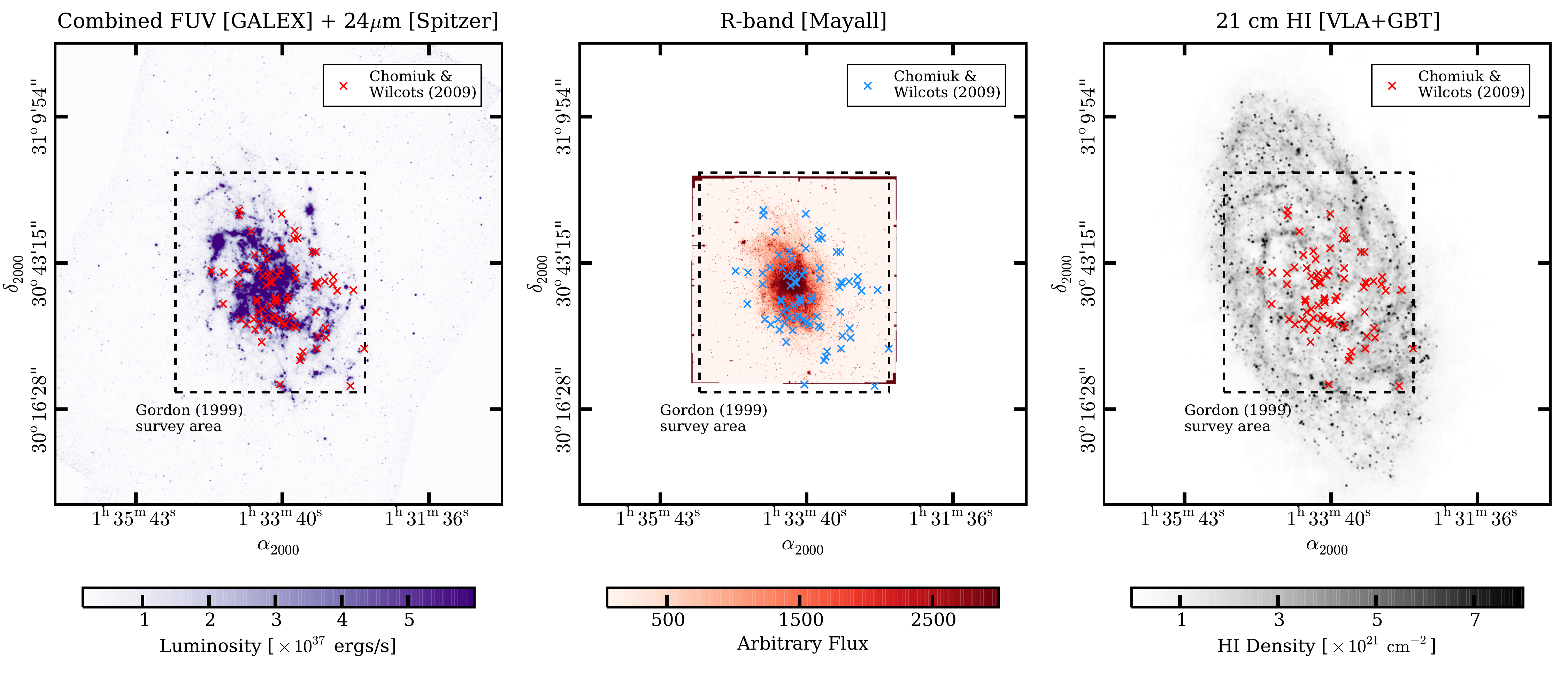}
\caption{Multi-wavelength maps of M33, with their properties and usage described in the text and summarized in Table \ref{multiwavtable}.  Each plot shows the 77 radio selected SNRs \protect\citep{Chomiuk2009a} as colored `x'. The dashed square region marks the 40 sq. arc-minutes area surveyed for radio SNRs by \protect\cite{Gordon1999}. All images were scaled to the resolution of 21 cm HI image.}
 \label{multiwavmap}
\end{figure*}

The spatial location of each SNR, depending on whether it is a Type Ia or CC, is selected from a set of multi-wavelength maps of M33 (Figure \ref{multiwavmap}, with details of the observations summarized in Table \ref{multiwavtable}). We assign locations of CC SNRs in M33 using FUV \protect\citep{GildePaz2007} and 24 $\mu$m \protect\citep{Dale2009} images, which trace recent star formation \protect\citep{Kennicutt2012}. Since FUV is prone to extinction, it is supplemented with the 24 $\mu$m image, which traces dust-enshrouded young stellar populations. We apply the correction in \protect\cite{Hao2011}, 
\begin{equation}
\rm{L_{FUV} = L_{FUV}^{obs} + 3.89\ L_{24 \mu m}}
\end{equation}
to produce the combined FUV and 24$\mu$m map for placing CC SNRs in our model. The probability of a CC SNR exploding in a given image pixel scales with the combined FUV and 24$\mu$m luminosity in that pixel. We chose FUV over the widely used H$\alpha$ line tracer of star formation, since H$\alpha$ reflects young stellar populations with ages up to 30 Myrs, whereas FUV emission mainly arises from young stellar populations with ages up to 100 Myrs, which accounts for the abundant, lower ZAMS mass progenitors of CC SNRs \protect\citep{Hao2011}. 

Type Ia SNRs are positioned by the distribution of stellar mass traced by an $r$-band image of M33 \protect\citep{Massey2006}. This is motivated by an observed correlation between the Type Ia rate with the $r$-band luminosity of the host galaxy \protect\citep{Yasuda2010}. Note that we only place SNRs in the survey area of \protect\cite{Gordon1999} shown by the dashed box in Figure \ref{multiwavmap}. This is to ensure comparison of our model within the same area that contains the observed SNRs. Although the actual survey area may be close to circular, a square survey area is not a bad approximation since very few model SNRs will be going off near the corners because of the low UV, IR and optical luminosities in these regions.

The vertical height $z$ at which the SNR goes off, for a given spatial location in M33, is drawn from an exponential distribution \protect\citep{Yoachim2006, Juric2008},
\begin{equation} \label{eq:pz}
p(z) \propto \rm{exp}\left(-\frac{|z|}{z_*}\right)
\end{equation}
where $z_*$ is the SN scale height. We set $z_* = 90$ pc for CC SNRs and 320 pc for Type Ia, assuming that CC SNe follow the distribution of bright stars near the Galactic plane and Type Ia SNe follow the bulk stellar disk population \protect\citep{Miller1979, Heiles1987}. These values are commonly assumed in simulations of SN-driven ISM \protect\citep[see][and references therein]{Joung2006, Hill2012a, Girichidis2016}. We comment on the effects of changing these assumptions in Section \ref{caveat}.\\

Our method of placing SNRs does not explicitly assume any particular form of DTD. We are simply stating that there are two kinds of SN progenitors - one associated with young, dense star forming environments, and one that is not correlated with recent star formation, according to observations. This simple approach is sufficient for our proof-of-concept study. More sophisticated placement strategies are certainly possible, and in the future we will explore them and the influence they can have on the derived DTD.

\subsubsection{Ambient Densities}
We assume a disk-like distribution of the M33 ISM and infer a volumetric density at the vertical position $z$ of the SNR, 
 \begin{equation} \label{n0z}
n_0(z) = \frac{N_H}{\sqrt{\pi z_0^2}} \mathrm{exp}\left(-\frac{z^2}{z_0^2}\right)
\end{equation}
where $N_H$ is the HI column density. We select $N_H$ at the spatial location of each SNR (selected in Section \ref{subsec:spat} ) from an HI column map corrected for self-absorption \protect\citep[shown in Figure \ref{multiwavmap}, ][]{Braun2012}. The scale height of the HI column, $z_0$ is treated as a free parameter in our model. More precisely, $z_0$ is the apparent scale height of M33, owing to its inclination. The Gaussian form of our ISM disk follows from observations of the vertical HI distribution near the Galactic mid-plane \protect\citep{Dickey1990}. 

The motivation for our relatively simple ISM model is to explore the parameter space of SNR evolution. Disk galaxies have a complex, multi-phase and inhomogeneous ISM \protect\citep{McKee1977, Hopkins2012, Hill2012a, Gent2013} that may modify the evolution of SNRs \protect\citep{Martizzi2015, Li2015}. But the assumption of a uniform ISM traced by HI serves as a good first approximation since HI has the highest ISM volume filling factor \protect\citep{Ferriere2001} and has been previously used to constrain the properties of various SNe and SNRs \protect\citep[e.g. ][]{Badenes2006, Chomiuk2012a, Chomiuk2016}. In a future paper, we will make use of the  analytical fits for radius and velocity of an SNR shock in an inhomogeneous, turbulent medium by \protect\cite{Martizzi2015} to understand the extent of modification imposed by an inhomogeneous ISM.

\subsubsection{Kinetic Energy and Ejecta Mass} \label{subsec:e51mej}
We draw kinetic energies, $E$ for Type Ia and CC SNe from a log-normal distribution centered on $10^{51}$ ergs,
\begin{equation}
p(\mathrm{log E}) = \frac{1}{\sqrt{2 \pi \sigma^2_{\mathrm{log E}}}} \exp{\left(-\frac{\left(\mathrm{logE} - \mu_{\mathrm{logE}}\right)^2}{2 \sigma^2_{\mathrm{logE}}}\right)}
\end{equation}
with $\mu_{\mathrm{logE}} = 51$ chosen for both Type Ia and CC SNe. The 1$\sigma$ error of the CC SN kinetic energy distribution, $\sigma_{\mathrm{logE}} = 0.28$, is chosen such that the fraction of normal CC (with $10^{51}$ ergs) to energetic gamma-ray bursts/hypernovae (with $\gtrsim 10^{52}$ ergs) is $\sim 10^{-3}$, consistent with observations \protect\citep{Podsiadlowski2004, Smartt2009}. Type Ia's, on the other hand, are a more homogenous class of explosions, and are expected to have a narrower spread in kinetic energies. We chose $\sigma_{\mathrm{logE}} = 0.1$, which is consistent with the observed rates of super-luminous and sub-luminous Type Ia \protect\citep{Meng2010} as well as the range of energies inferred from the X-ray spectra of well studied Type Ia SNRs \protect\citep{Badenes2008}. We will discuss the effect of changing our assumptions on the kinetic energy distributions of Type Ia and CC SNe in Section \ref{caveat}.

Even though SNe can have different ejecta masses, we fix the ejecta masses $\rm{(M_{ej})}$ for CC SNRs and Ia's at $5\ \rm{M_{\odot}}$ and $1.4\ \rm{M_{\odot}}$ respectively. This is justified because SNR surveys are dominated by objects in the Sedov stage of evolution where ejecta mass has a negligible effect on the evolution (See Section \ref{subsec:rlc} and Figure \ref{fig:n0em}).

\subsubsection{Radio Light Curve} \label{subsec:rlc}
\begin{figure}
\includegraphics[width=\columnwidth]{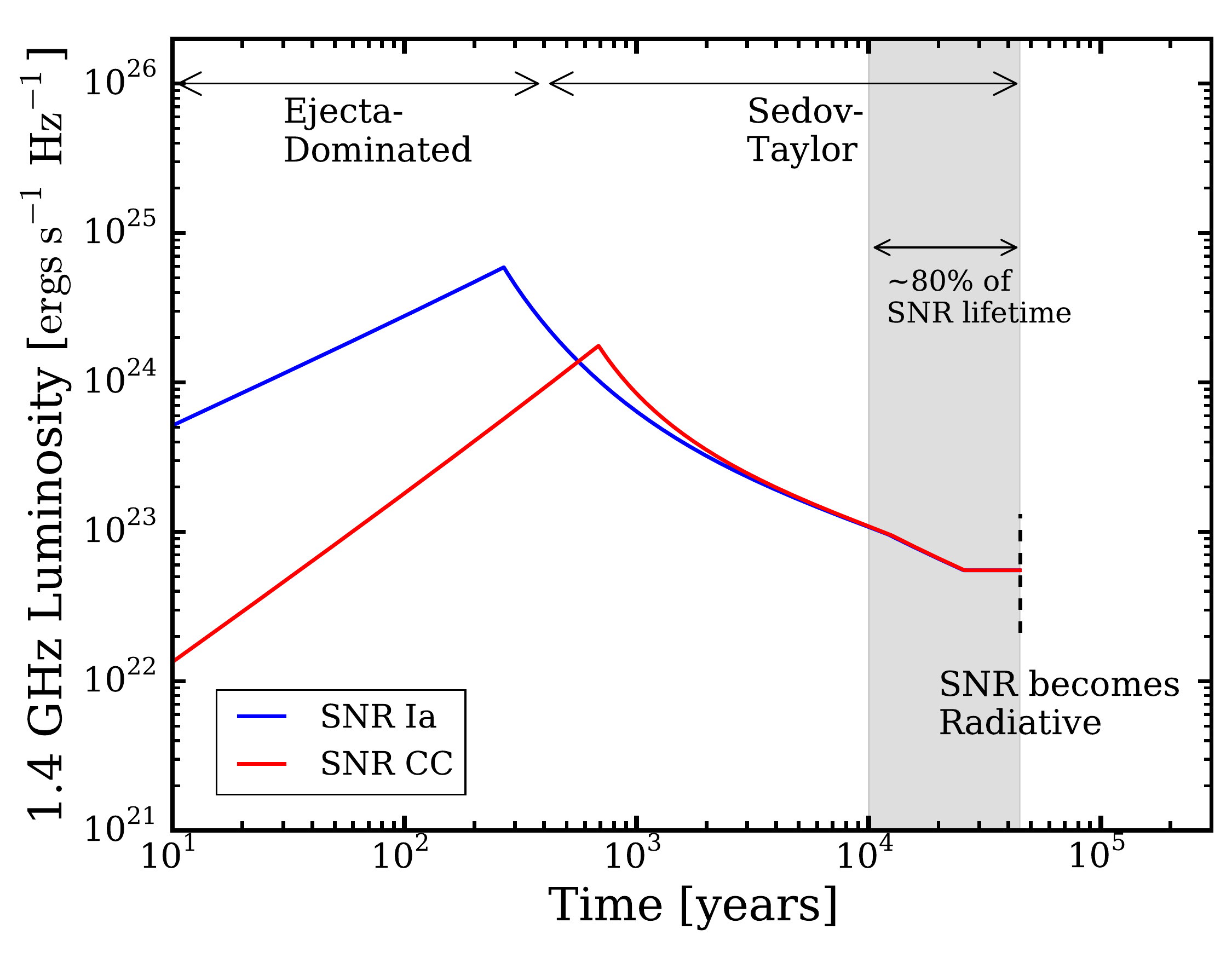}
\caption{Illustration of a model radio light curve for Type Ia and CC SNR, evolving through the ejecta-dominated and Sedov-Taylor phases. The dashed line marks the transition to the radiative phase, where we assume the synchrotron emission becomes inefficient, as discussed in Section \ref{vistime}. Both types of SNRs exploded with kinetic energy of $10^{51}$ ergs in an ISM density, $n_0 = 1$ $\rm{cm}^{-3}$. For CC, we chose $M_{\rm{ej}} = 5$ $\rm{M_{\odot}}$, and for Ia, $M_{\rm{ej}} = 1.4$ $\rm{M_{\odot}}$. Most SNRs will be found in the Sedov-Taylor stage, as shown by the shaded region.}
\label{fig:radiolightcurve}
\end{figure}

\begin{figure*}
\includegraphics[width=\textwidth]{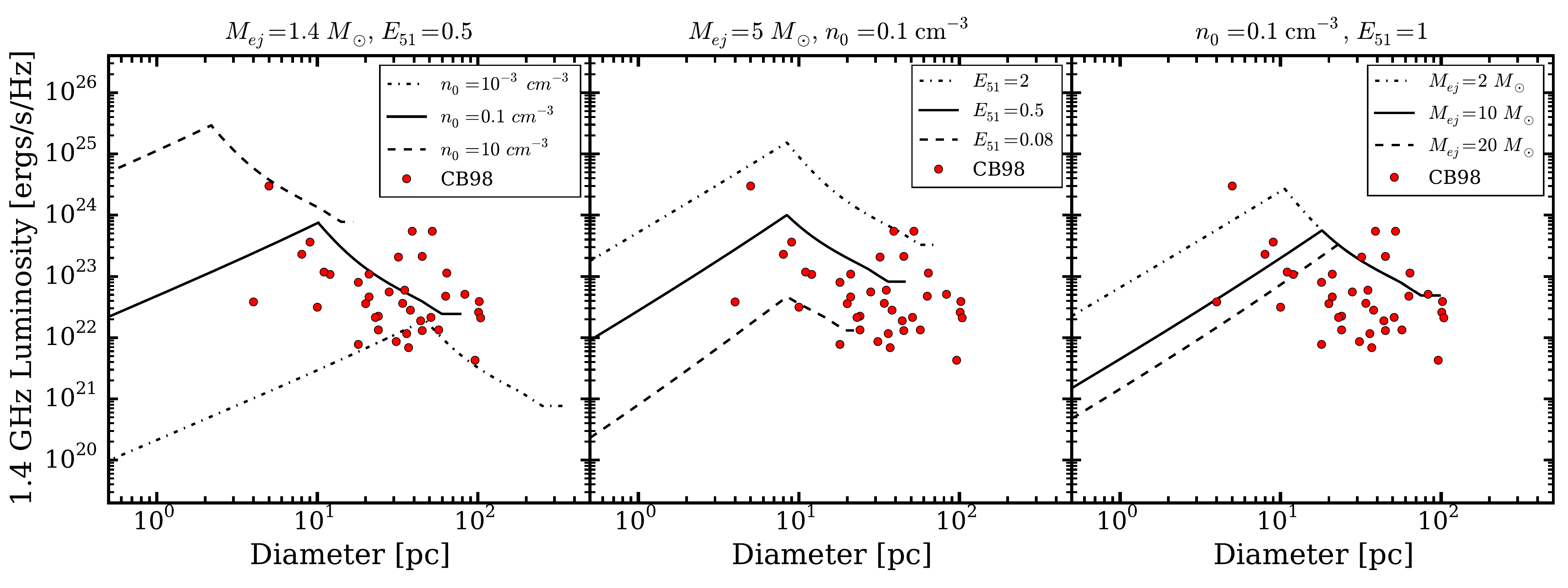}
\caption{The effects of varying the ISM density ($\rm{n_0}$), kinetic energy ($E_{51}$, in units of $10^{51}$  ergs) and ejecta mass ($\rm{M_{ej}}$, in units of $\rm{M_{\odot}}$) on the SNR light curves. For each panel, we fix $\epsilon_{\rm{e}} = 5 \times 10^{-3}$, and hold the parameters in the title constant, while changing the parameter in the legend box. E.g. in the left panel, we show radio light curves for varying $n_0$, at fixed $M_{ej}$ and $E_{51}$. The red data points are SNRs taken from \protect\cite{Case1998}, denoted as CB98.}
\label{fig:n0em}
\end{figure*}

For each SNR generated, we calculate its radio luminosity at a given age with a synthetic radio light curve model. Radio emission in SNRs comes from synchrotron radiation emitted by electrons that are accelerated to relativistic energies by the shock \protect\citep{Bell1978, Chevalier1982a, Chevalier1998, Berezhko2004a, Chevalier2006}. The accelerated particles produce streaming instabilities ahead of the shock, which strongly amplifies the magnetic field \protect\citep{Bell2004, Amato2009, Caprioli2014a} that further contributes to the radio emission \protect\citep{Thompson2009, Chomiuk2009a}. 

In this section, we will only explain the main features of our SNR light curve model. The detailed derivation of the radio luminosity, based on theories of diffusive shock acceleration, field amplification and shock dynamics is described in Appendix \ref{app:rlc}. The radio luminosity of SNRs at 1.4 GHz is optically thin synchrotron emission given by, 
 \begin{equation}
\begin{split}
L_{1.4}\ \approx\ & (2.2 \times 10^{24}\ \mathrm{ergs/s/Hz}) \\
& \left(\frac{R_{\rm{s}}}{10\ \mathrm{pc}}\right)^3 \left(\frac{\epsilon_{\rm{e}}}{10^{-2}}\right) \left(\frac{\epsilon^{\mathrm{u}}_{\mathrm{b}}}{10^{-2}}\right)^{0.8} \left(\frac{v_{\rm{s}}}{500\ \rm{km/s}}\right)^{3.2} \\
\end{split}
\end{equation}
\begin{figure}
\includegraphics[width=\columnwidth]{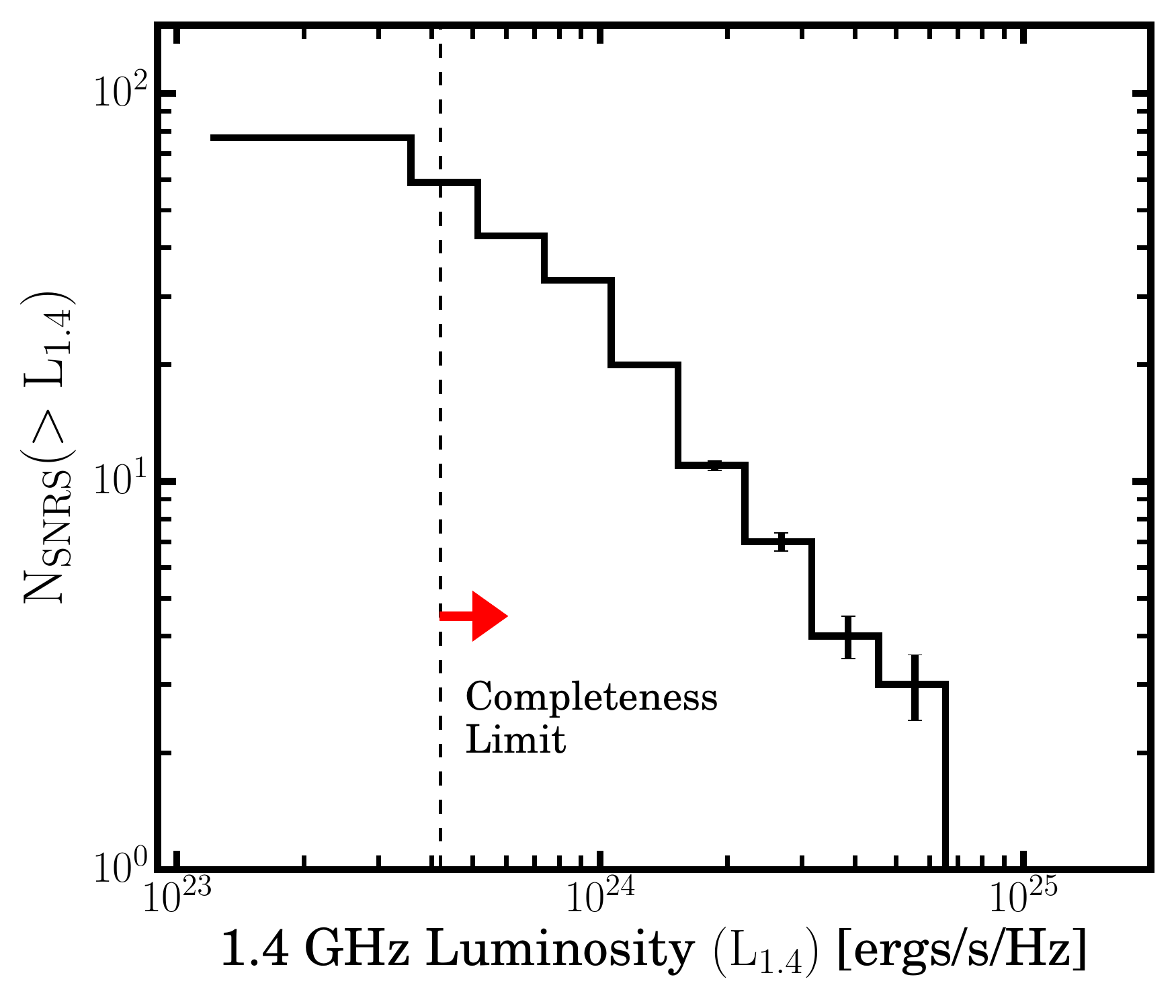}
\caption{The cumulative LF of 77 SNRs in M33 \protect\citep{Chomiuk2009a}, with Poissonian error bars, used for constraining the model parameters, $R$, $z_0$ and $\epsilon_{\rm{e}}$. The completeness limit is indicated by the dashed vertical line at $4.2 \times 10^{23}$ ergs/s/Hz, and 59 SNRs that have luminosities greater than this limit are considered part of the `complete' SNR survey. }
\label{fig:obsLF}
\end{figure}
where $R_s$ is the shock radius and $v_s$ is the shock velocity. We assume a small fraction $\epsilon_{\rm{e}}$, of the shock energy is shared by the relativistic electrons accelerated by the shock (referred to as electron acceleration efficiency in this paper) and is considered a free-parameter in our model. The spectrum of the acceleration electrons is $N(E) = N_0 E^{-p}$ based on the theory of diffusive shock acceleration. We fix $p$ = 2.2 from gamma-ray observations of Galactic SNRs \protect\citep{Morlino2012, Caprioli2012} and in Section \ref{caveat}, we discuss the effects of changing our assumptions on $p$. 

A novel aspect of our light curve model is the treatment of magnetic field amplification, which is induced upstream of the shock \protect\citep{Morlino2010, Caprioli2014a} and parameterized by $\rm{\epsilon_{b}^{u}}$, the fraction of shock energy contained in the amplified upstream magnetic field, $B_u$. The downstream field, $B$ is then a simple compression of $B_u$. Instead of leaving $\rm{\epsilon_{b}^{u}}$ as a free parameter, we use results of simulations of particle acceleration from first principles \protect\citep{Caprioli2014a} to scale $\epsilon^{\mathrm{u}}_{\mathrm{b}}$ with the Alfv\'{e}n Mach number of the shock and the cosmic ray acceleration efficiency (Eq \ref{eq:epsu}).  The form of the scaling depends on whether the amplification is induced by resonant \protect\citep{Bell1978} or non-resonant \protect\citep{Bell2004} streaming instabilities (see Appendix \ref{app:mfa} for a detailed discussion). No equipartition between energy densities of the magnetic fields and relativistic electrons is assumed.

\begin{figure*} 
\includegraphics[width=0.8\textwidth]{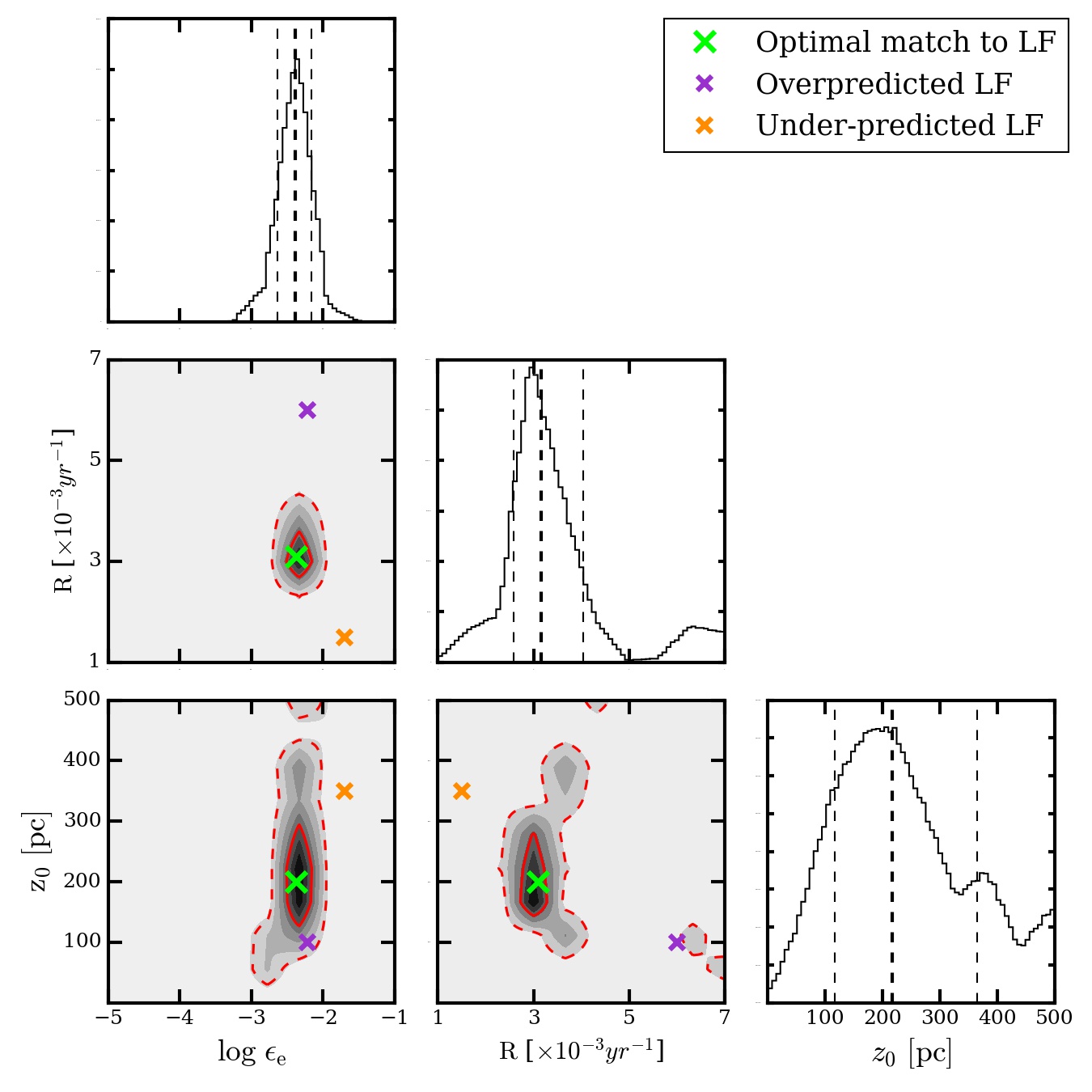}
\caption{The parameter space of our model constrained by the observed SNR LF in M33 in Figure \ref{fig:obsLF}. The dark contours represent favorable regions of parameter space where the model reproduces the observed SNR LF, with solid and dashed red lines showing the 1$\sigma$ and 2$\sigma$ contours respectively. The histograms are the marginal probability densities of each parameter, with bold dashed line being the median, and the normal dashed lines showing the 16th and 84th percentiles. As a consistency check, we pick values of $R, z_0$ and $\epsilon_{\rm{e}}$ from different areas of parameter space (shown by colored crosses) and compare the predicted LFs in Figure \ref{fig:compare}.}
\label{fig:paramspace}
\end{figure*}
The SNR radio light curve (Figure \ref{fig:radiolightcurve}) is determined by the competing effects of expansion, which increases the emitting volume ($\sim R_s^3$), and deceleration (decreasing $v_s$), which reduces the abundance of relativistic electrons and the streaming instabilities that amplify the magnetic field. Expressions for $R_s$ and $v_s$ as the SNR evolves from ejecta-dominated to Sedov-Taylor phase are given in Table \ref{table:rs}. During the ejecta-dominated phase (the first $10^2 - 10^{3}$ years), $L_{1.4}$ increases as the shock expands rapidly through the ISM. The light curve peaks at the beginning of the Sedov-Taylor phase, once the shock has swept-up an ISM mass equivalent to the ejected mass. At this point, a period of enhanced deceleration ($v_s \propto t^{-3/5}$) slows down the expansion and causes $L_{1.4}$ to drop. When $B$ ($\propto v_s$) drops to a simple compression of the ambient magnetic field, $B_0$ (Eq. \ref{eq:B0}), we fix $B = 4 B_0$, which balances the effects of expansion and deceleration, causing the light curve to flatten out. The general shape of our analytical light curve is consistent with numerical calculations \protect\citep{Berezhko2004a}. The slight curvature around $10^3$ years represents the increasing contribution from resonant instabilities, and the `knee' (around $10^4$ years) is because of the decreasing particle acceleration efficiency (Appendix \ref{app:mfa}). Note that the SNR spends most of its lifetime in the Sedov-Taylor phase, and therefore \emph{most Local Group SNRs are expected to be found in their Sedov phase}.

The SNR radio light curve also depends on the ambient medium density, kinetic energy and ejecta mass of each SNR (Figure \ref{fig:n0em}). Denser ISM causes SNRs to decelerate faster (hence, smaller diameters), and remain radio visible for a shorter period, assuming they remain above the survey detection limit. SNRs in denser ISM are also brighter due to greater energy available to the downstream magnetic field and relativistic electrons (Eq \ref{N0}, \ref{eq:bu}). In the same way, higher kinetic energies also produce brighter SNRs, but these SNRs are larger and radio visible for longer periods because of greater energy in the forward shock.
Finally, the ejecta mass mainly affects SNRs in their ejecta-dominated phase, with higher ejecta masses implying smaller energy per unit mass of ejecta (assuming energy is conserved) and thus fainter SNRs. The onset of the Sedov-Taylor phase is also delayed because the SNRs need to sweep up a larger mass of ISM (equal to their ejected mass). However, \emph{the ejecta mass does not affect the Sedov-Taylor light curve}, since the evolution in this phase is driven by the swept-up ISM, which far exceeds the initial ejecta mass. Because of this, we kept the ejecta masses of our model Type Ia and CC SNRs fixed in Section \ref{subsec:e51mej}. 

We also show in Figure \ref{fig:n0em} that for reasonable ranges of values of ISM densities, kinetic energies and ejecta masses, our model predicts radio luminosities and diameters that are similar to well-known SNRs in the Galaxy \protect\citep{Case1998}. This is a consistency check. The galactic SNRs are not used to constrain the model light curves, since reproducing the individual luminosities and diameters require more detailed modeling of the SNR and its environment.

\subsection{Comparison with SNR catalogs}
Within the assumptions discussed in Section \ref{genSNRs}, our model has three free parameters - SN Rate ($R$), HI scale height ($z_0$) and electron acceleration efficiency ($\epsilon_{\rm{e}}$). We can constrain these parameters by comparing the radio LF of our steady-state model SNR populations with the observed SNR LF in M33. 

Figure \ref{fig:obsLF} shows the radio LF of SNRs in M33 from the catalog of \protect\cite{Chomiuk2009a}, and whose locations are also shown in Figure \ref{multiwavmap}. The catalog consists of 77 SNRs, a subset of the 186 radio sources in M33 compiled by \protect\cite{Gordon1999} from 6 and 20 cm observations \protect\citep{Duric1993} having a noise limit of 50$\mu$Jy per beam, where the beam diameter was 7" or 29 pc at 840 kpc, the adopted distance to M33. The 77 SNRs were classified by their radio images (sources with 3$\sigma$ detection above noise, with a synchrotron signature $S_{\nu} \propto \nu^{-\alpha}$, where $\alpha \geq 0.2$ was used to distinguish from HII regions) and by detection of an H$\alpha$ counterpart to distinguish them from background galaxies, which would be redshifted out of the narrowband filter. 

Because of the homogenous selection criteria for our SNR catalog, we are able to define a \emph{completeness limit} above which we consider the sample to be complete. SNRs can be missed if they are below the noise sensitivity of the radio survey, which increases at larger radii from the phase center of the image and in  regions of vigorous star-formation. To account for these SNRs missing from the faint end of the LF, \protect\cite{Chomiuk2009a} showed that the radio LFs of SNRs can be adequately described as a power-law, and defined the completeness limit of an SNR sample as the flux where the LF `turns over'. We set the completeness limit of M33 SNRs at $4.2 \times 10^{23}$ ergs/s/Hz, and 59 of the 77 SNRs have luminosities above this limit.
 
We assumed flat priors for our model parameters, $R$, $z_0$ and $\epsilon_{\rm{e}}$. The star-formation rate of M33 is about a factor of 10 lower than that of the Milky Way \protect\protect\citep{Chomiuk2011}, so we assumed the M33 SN rate is similarly lower than that of a Milky Way-like spiral \protect\protect\citep{Li2011} and set the prior for $R$ within $(1 - 7) \times 10^{-3}$ per year. We set $z_0$ within 0 - 500 pc, which is the net extent of the neutral phase seen in most SN-driven ISM simulations \protect\protect\citep{Hill2012a, Gent2013, Walch2015, Kim2015}. The electron acceleration efficiency is assigned a wide prior in logarithmic space, $10^{-1} - 10^{-5}$. The upper and lower limits for the prior on $\epsilon_{\rm{e}}$ are roughly consistent with values derived for radio SNe and young SNRs respectively (see Section \ref{subsec:epse}). The statistical comparison of the model and observed radio LFs is done with a maximum likelihood method described in Appendix \ref{app:maxlik}.
\section{SNR model parameter space} \label{sec:PARAM}
\label{sec:paramspace}
\begin{figure} 
\includegraphics[width=\columnwidth]{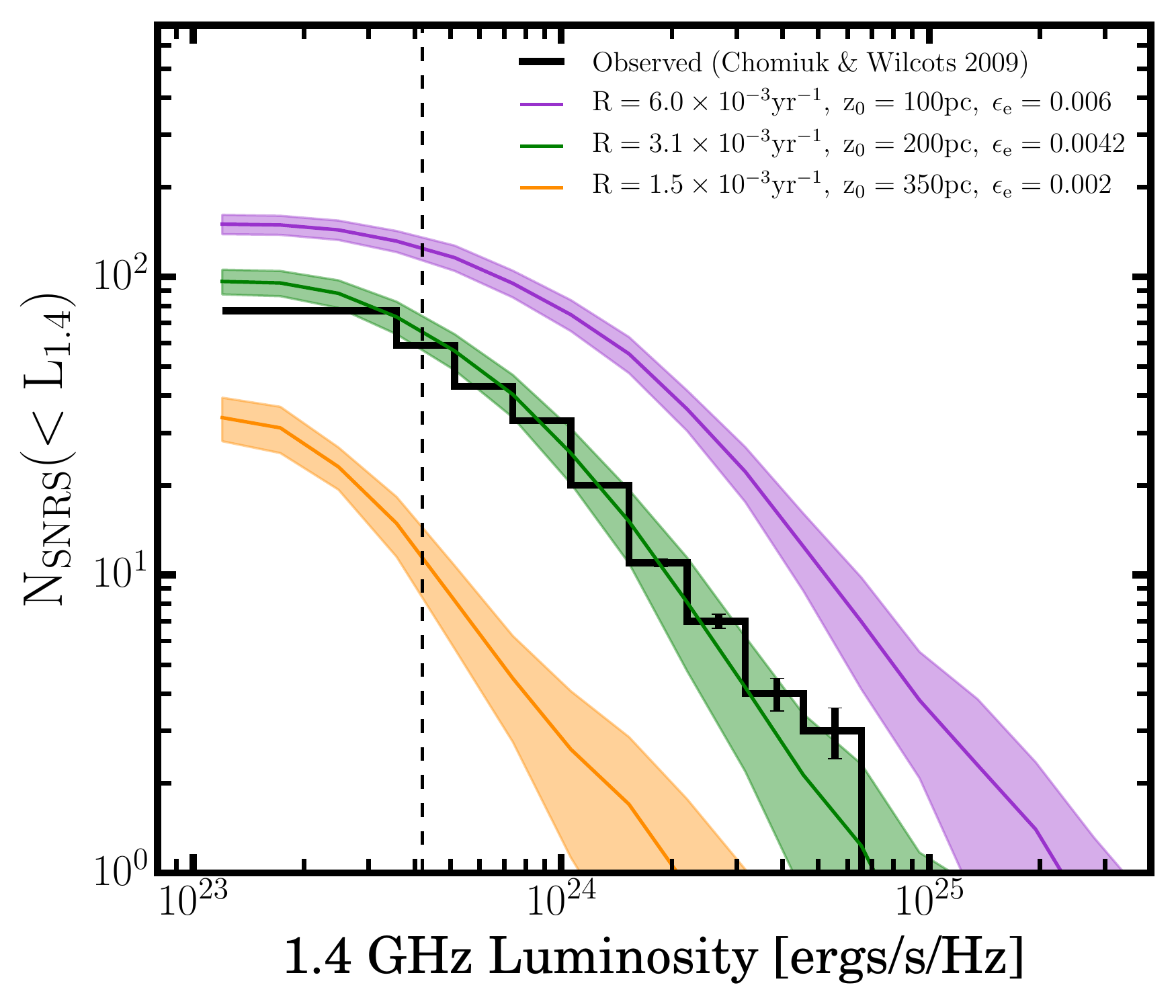}
\caption{Comparison of the observed M33 SNR LF (black) with different LFs (colored) predicted by the model. Green LF corresponds to the best fits values of $R, z_0$ and $\epsilon_{e}$ (green crosses in Figure \ref{fig:paramspace}) that reproduce the observed LF, while the purple and orange LFs (purple and orange crosses in Figure \ref{fig:paramspace}) are inconsistent with observations. The colored solid lines and shaded region represent the median and $\pm 1\sigma$ uncertainties of the LFs respectively. The vertical dashed line is the completeness limit = $4.2 \times 10^{23}$ ergs/s/Hz. }
\label{fig:compare}
\end{figure}
 \begin{figure}
\centering
\includegraphics[width=\columnwidth]{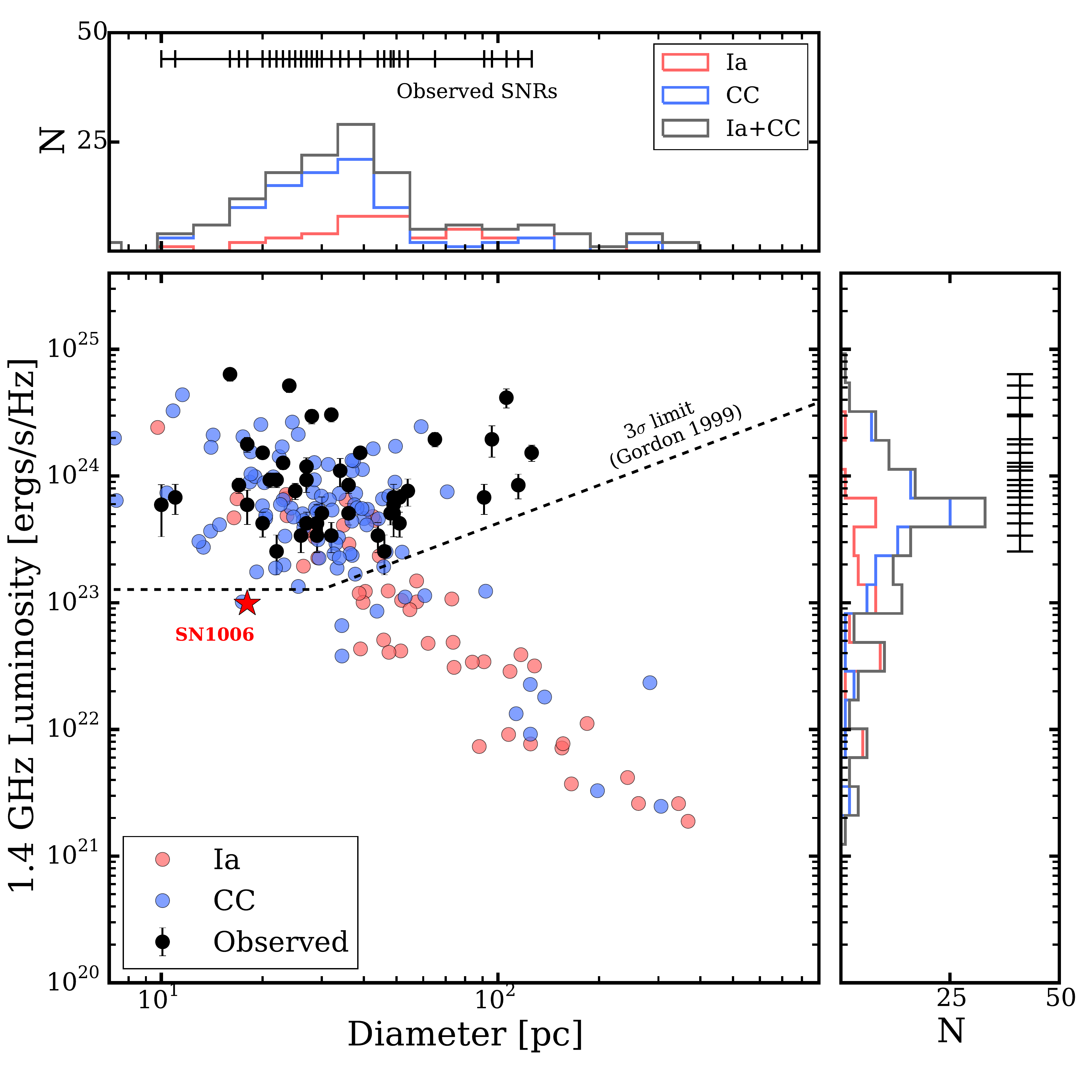}
\caption{\textit{Central panel} shows a snapshot of a best-fit steady state model population produced for $R = 3.1 \times 10^{-3}$ SN per year, $z_0 = 200$ pc and $\epsilon_{\rm{e}} = 4.2 \times 10^{-3}$ in luminosity-diameter space. The red and blue symbols represent model Type Ia and CC SNRs respectively. The solid data points are 44 of the 77 radio-selected SNRs \protect\protect\citep{Chomiuk2009a} that have diameters available from optical images \protect\protect\citep{Gordon1998}. The 3$\sigma$ detection limit for the \protect\cite{Gordon1999} survey ($\sigma = 50 \mu$Jy/beam with a 7 sq. arc-seconds beam at 840 kpc, the distance to M33) is shown with a dashed black line. \textit{Upper panel} shows the diameter histograms of the model SNRs, compared to the observed 44 SNRs shown with black ticks. \textit{Right panel} shows the same information, but for the 1.4 GHz luminosity.}
\label{fig:lumdiam}
\end{figure}
The model parameter space constrained by the M33 SNR catalog is shown in Figure \ref{fig:paramspace}. Darker areas represent parameter values for which our model and observations agree, while the contrary is true for lighter areas. Visual comparisons between the predicted and observed LFs are shown in Figure \ref{fig:compare} for parameter values taken from different regions in Figure \ref{fig:paramspace}. Values of $R, z_0$ and $\epsilon_{\rm{e}}$ taken from the darkest regions predict LFs that agree with the observed LF above the completeness limit, whereas the ones from lighter regions do not. The discrepancy with observations below the completeness limit for the green histogram implies that faint SNRs may be missing from the catalog, and this issue is discussed further in Section \ref{sec:misssnr}. The small discrepancy at the bright end ($L_{1.4} > 3 \times 10^{24}$ ergs/s/Hz) is because the statistical comparison between the model and observed LFs is dominated by SNRs near the completeness limit, which are larger in number. Nonetheless, the green LF agrees within the Poisson errors of the bright end of the observed LF.

Figures \ref{fig:paramspace} and \ref{fig:compare} highlight the crucial role of SNR catalogs in our model. We can use them to constrain the unknown parameters in our model, and then use their observationally-constrained values to predict accurate radio visibility times (Section \ref{vistime}). The radio LF of SNRs rules out significant areas of parameter space for parameters of physical interest, such as $R$ and $\epsilon_{\rm{e}}$. For M33, we measure a median value of $R \sim 3.1 \times 10^{-3}$ SN per year and $\rm{log}\ \epsilon_{\rm{e}} \approx -2.38$, or $\epsilon_{\rm{e}} \approx 4.2 \times 10^{-3}$. Since the radio LFs are particularly sensitive to $\epsilon_{\rm{e}}$ ( $L_{1.4} \propto \epsilon_{\rm{e}}$), we obtain tighter constraints on $\epsilon_{\rm{e}}$ than the other parameters. For higher values of $\epsilon_{\rm{e}}$, SNRs are more luminous and this shifts the bright end of the cumulative LF right, towards higher luminosities. While $R$ does not affect the radio luminosities of individual SNRs, it directly controls the number of SNRs per luminosity bin. Therefore, increasing $R$ shifts the LF up (towards higher $\rm{N_{SNR}}$), with all other parameters fixed. The HI scale height $z_0$ is the least sensitive of the three parameters and has the poorest constraints. This is because we choose the vertical heights of SNRs in the ISM ($z$) from the stellar distribution (Eq. \ref{eq:pz}), which produces SNRs in a wide range of $z$ ($z_* = 90$ pc for CC SNRs and $320$ pc for Type Ia SNRs). Thus, changes in the value of $z_0$ within this range do not affect the radio LFs of the model SNRs as much as $R$ or $\epsilon_{\rm{e}}$. However, the constrained values of $z_0$ still fall intermediate to the assumed scale heights for the young and old stellar populations. Although these measurements are based on the partial 40 sq. arc-minute coverage of the M33 disk in Figure \ref{multiwavmap}, they can be scaled to the entire galaxy with a SNR survey that covers the entire disk. We only show the median values of our measurements in Figure \ref{fig:paramspace} because of the asymmetric nature of the probability histograms (particularly $z_0$), and discuss the effects of changes in the parameter space in Section \ref{caveat}.
\begin{figure*}
\includegraphics[width=0.7\textwidth]{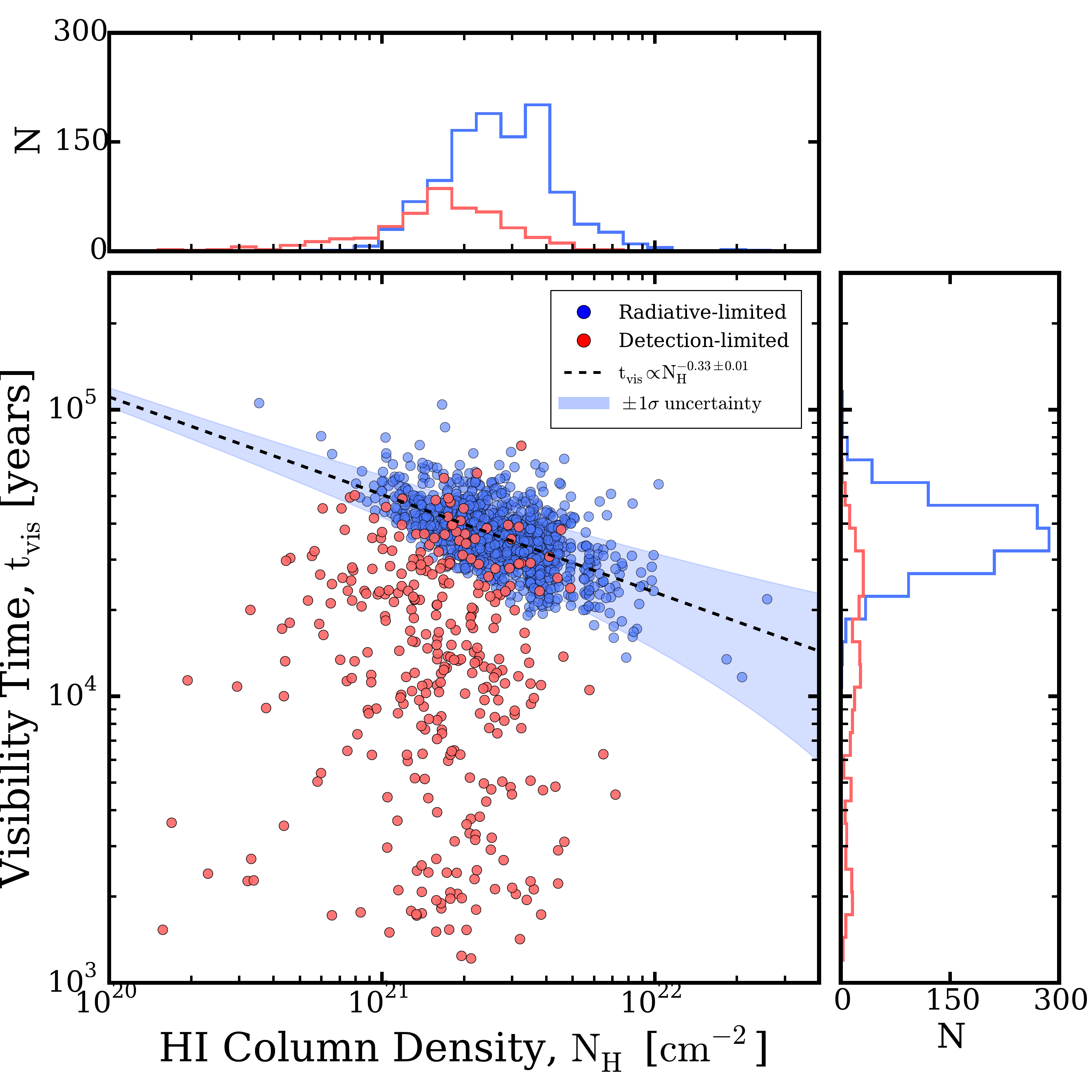}
\caption{Radio visibility times of SNRs generated for $R = 3.1 \times 10^{-3}$ SN per year, $z_0 = 200$ pc and $\epsilon_{\rm{e}} = 4.2 \times 10^{-3}$. Blue circles represent visibility times determined by the transition to the radiative phase, while red points represent visibility times determined by the 3$\sigma$ detection limit of the \protect\cite{Gordon1999} survey. The same color scheme is followed for the histograms for column density and visibility times in the upper and right panels respectively. The line of regression fit to the radiative-limited population is shown with a dashed line, with the shaded region representing $\pm 1\sigma$ interval}.
\label{fig:vistime}
\end{figure*} 

The predicted and observed size distribution of SNRs in M33 are also consistent as shown in Figure \ref{fig:lumdiam}, even though we did not attempt to fit the SNR sizes. However, the model does not reproduce the largest SNRs in M33 (> 80 pc), which we discuss later in Section \ref{sec:misssnr}. A conspicuous feature is the correlation between the luminosities and diameters of SNRs over several decades, which is not apparent in the observed SNRs. This is because large SNRs are usually expanding into low ambient densities, and thus experience weaker electron acceleration and field amplification, yielding smaller radio luminosities. For a given size, the scatter in radio luminosities comes from the range of kinetic energies and ambient densities of SNRs. Type Ia's show a smaller scatter than CC's in the model because Type Ia's are drawn from a narrower range of kinetic energies. We note that SNRs exploding in densities $< 10^{-3}$ $\rm{cm^{-3}}$ do not appear in our model since these are densities more characteristic of the warm ionized phase of the ISM \protect\citep{Draine2011}. 

Our model also predicts that radio surveys will be dominated by CC SNRs above the 3$\sigma$ detection limit, with characteristically higher luminosities and smaller diameters than Type Ia, since CC SNe preferentially evolve in higher ambient densities than Type Ia. For the best fit parameters, the fraction of Ia/CC SNRs produced above the 3$\sigma$ detection limit of the \protect\cite{Gordon1999} survey is 0.1-0.25, less than our input fraction of Type Ia/CC SNe of 1/3. Because of the characteristically different light curves of Type Ia and CC SNRs, the ratio of Ia/CC SNe cannot be estimated from SNR surveys that do not account for completeness limits.

\section{SNR Visibility Time} \label{vistime}
Deriving the visibility times of SNRs is more challenging than SNe. Firstly, the ages of SNRs are difficult to estimate. Only SNRs with historical records or light echoes have reliably determined ages, and most of these are young SNRs. Secondly, we do not have observations of SNR light curves on their characteristic timescales ($\sim 10^4$ years). Thirdly, reliable classification of SNRs as Type Ia or CC in origin is not possible except for young SNRs using e.g. X-ray spectra \protect\citep{Badenes2010a, Yamaguchi2014}, light echoes \protect\citep{Rest2005} or association with neutron star or pulsars. Older SNRs are much harder to classify as their spectra is dominated by the swept-up ISM, although their origins may be indirectly guessed from morphology \protect\citep{Lopez2009, Lopez2011,Peters2013} or association with stellar populations \protect\citep{Badenes2009, Lee2014, Lee2014a, Maggi2016}. Most of these techniques become increasingly difficult for distances beyond the Magellanic Clouds. 

As an alternative, \protect\cite{Badenes2010} estimated the visibility time of SNRs in a given subregion of the Magellanic Clouds from the \emph{in-situ} ISM column density. SNRs in regions of denser ISM will be brighter but remain visible for shorter periods of time than SNRs in tenuous ISM. Assuming all Magellanic SNRs are in their Sedov stages and that the SNR sample was complete, the visibility time was approximated as $t_{\rm{vis}} \propto N_H^{-1/2}$. This is related to the cooling timescale of a Sedov SNR transitioning to the radiative phase, with a cooling function of the form, $\Lambda(T) \propto T^{\epsilon}$ for $10^6$ K shocked gas, with $\epsilon = -1/2$ to -3/2. 

In this paper, the visibility time of each SNR can be derived from its radio light curve (Figure \ref{fig:radiolightcurve}), which is terminated when either of these two conditions are met: (1) If the SNR falls below the detection limit of the radio survey. For the M33 catalog, the detection limit is 3$\sigma$ above the rms sensitivity of 50$\mu$Jy per beam for the \protect\cite{Gordon1999} survey. (2) If the SNR shock transitions to the radiative phase  \protect\citep{Bandiera2010}. At this point, we assume the synchrotron mechanism becomes inefficient because the magnetic field amplification is too low, and not enough relativistic electrons are produced by the weakened SNR shock \protect\citep{Berezhko2004a, Blasi2007}. To the first order, we assume this radiative transition occurs when $v_s \approx 200$ km/s \protect\citep{Blondin1998}.

Figure \ref{fig:vistime} shows the relation between the SNR radio visibility times and the ambient HI column density for a population of model SNRs generated using the best-fit parameter values (from Figure \ref{fig:paramspace}). Most SNRs have visibility times between 20-80 kyrs, and correlated with the column density as $t_{\rm{vis}} \propto N_H^{-a}$ with $a = 0.33 \pm 0.01$. SNRs in high column densities evolve in higher ambient densities and so, being brighter, remain above the detection limit. As a result, their radio visibility is decided by their transition to the radiative phase (when $v_s \approx$ 200 km/s). This occurs quicker at higher column densities because of the strong deceleration by the dense ISM. The scatter in the relationship is due to SNR light curves with different ambient densities and kinetic energies. Most CC SNRs will have visibility times decided by the transition to the radiative phase, since they explode in higher densities and have surface brightnesses above the detection limit of the radio survey. About 30$\%$ of the SNRs, mostly Type Ia exploding in lower densities, deviate from this relation by exhibiting smaller visibility times. These SNRs have low surface-brightness because of lower ambient densities, causing their light curves to be determined by the detection limit of the survey. The visibility times predicted by our model are consistent with the ages of SNRs associated with pulsars and magnetars \protect\citep{Martin2014}. These predictions can be further tested by future radio SNR surveys using images with deeper sensitivity limits (Huizenga et. al, in prep).

We therefore have a unique pathway to calculating the visibility times of an SNR population using a model that is equipped with the physics of SNR evolution and synchrotron emission, and observationally constrained by multi-wavelength surveys of the host galaxy. While our results nicely confirm the conjecture of \protect\cite{Badenes2010} that the visibility time of SNRs in the Local Group is roughly their Sedov-Taylor lifetimes, our model also has the added advantage of being able to characterize the population of detection-limited SNRs, which is significant at distances farther than the Magellanic Clouds. In addition, the model also helps us understand the scatter in visibility times arising from the diversity of ambient densities and kinetic energies of SNRs. Understanding these subtleties is crucial because the model-based visibility times will be the main source of systematic uncertainties in the derivation of a SN DTD in the Local Group.

\section{Correlation between the brightest SNR and the SN rate} \label{sec:BRIGHT}
\begin{figure}
\includegraphics[width=\columnwidth]{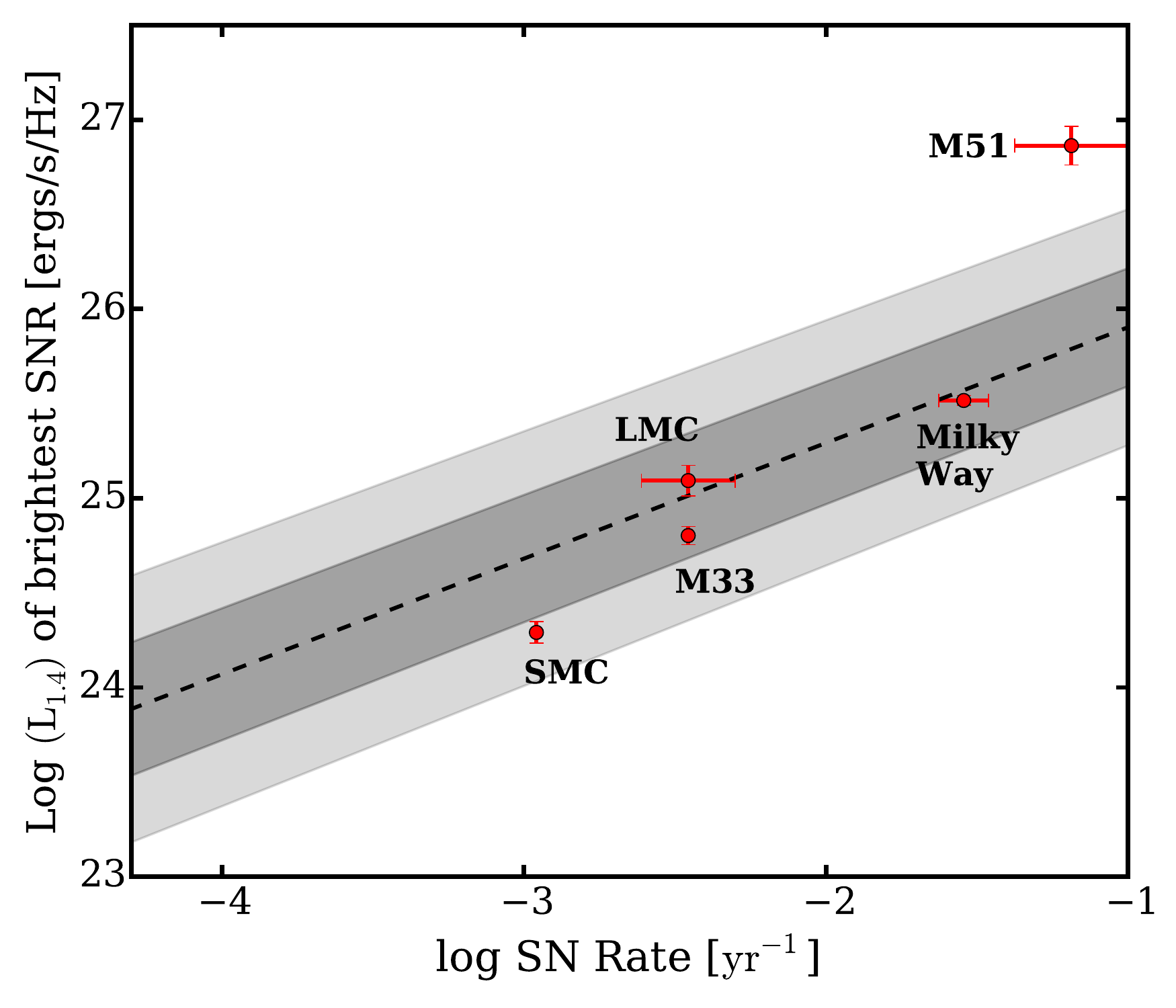}
\caption{Radio luminosity of the brightest SNR produced by the model as a function of SN rate, assuming $z_0 = 200$ pc and $\epsilon_{\rm{e}} = 4.2 \times 10^{-3}$. The two shades of grey represent $\rm{\pm1\sigma}$ and $\rm{\pm2\sigma}$ regions for our model. The red points represent galaxies with known SN rates (M51: \protect\cite{Rampadarath2015}, Milky Way: \protect\cite{Li2011a}, M33: this work, LMC: \protect\cite{Badenes2010a}, SMC: \protect\cite{Tammann1994}) and luminosities of the brightest SNRs in these galaxies \protect\citep{Chomiuk2009a}. }
\label{fig:brightsnr}
\end{figure}
We examine the correlation between the brightest model SNR and the SN rate in M33 as a consistency check for our model. \protect\cite{Chomiuk2009a} showed that host galaxies with the highest star formation rates (SFRs) produce the brightest SNRs. Such a trend can also be expected with SN rate, since it is proportional to the SFR. Figure \ref{fig:brightsnr} shows such a correlation exists for over two decades in radio luminosity, and 3 decades in SN rates in our model, assuming $z_0 = 200$ pc and $\epsilon_{\rm{e}} = 4.2 \times 10^{-3}$ as determined in Figure \ref{fig:paramspace}. The shaded region represents the spread in luminosities for a given SN rate due to different kinetic energies and ambient densities of the SNRs. For higher SN rates, there's a greater chance of a bright SNR exploding with a higher kinetic energy at explosion, making them more radio bright. The higher SN rate also increases the probability of the brightest SNR going off in a denser region of the M33 ISM.  A caveat with this analysis is that our model ISM remains unchanged with SN rate, whereas galaxies with higher SN rates will have a denser ISM \protect\citep{KennicuttJr.1998} and distinctly different ISM properties due to feedback \protect\citep{Hopkins2012}.

For comparison, we show the SN rates and radio luminosities of the brightest SNRs in nearby galaxies. Galaxies with higher SN rates host brighter SNRs, and with the exception of M51, falls within the $2\sigma$ shaded region. The discrepancy could be because the SFR of M51 is nearly twice that of the Milky Way \protect\citep{Chomiuk2009a} and forms an interacting pair with M51b. Galaxies with such high SFRs or extreme environments may also host objects that mimic spectral signatures of radio SNRs, such as supernovae, super-bubbles, remnants of hypernovae or ultra-luminous X-ray sources (ULXs) \protect\citep{Chomiuk2011}.\\\\\\\\
\section{Discussion} \label{sec:DISCUSS}
\subsection{Understanding electron acceleration in interstellar shocks with an SNR catalog} \label{subsec:epse}
We estimated the electron acceleration efficiency in SNR shocks, $\epsilon_{\rm{e}}$ from a catalog of SNRs mostly in their Sedov phase. On average, the M33 SNRs have $\epsilon_{\rm{e}} \sim 4.2 \times 10^{-3}$, which falls in between conventional estimates for radio SNe and young SNRs. Type Ib/c SNe with relativistic shocks commonly require $\epsilon_{\rm{e}} \sim 0.1$ to explain their radio light curves assuming equipartition between magnetic field and electron energy densities, or $\epsilon_{\rm{e}}/\epsilon^{\mathrm{u}}_{\mathrm{b}} = \alpha$ where $\alpha$ is a constant. \protect\cite[e.g. ][]{Chevalier1998, Li1999, Berger2002, Soderberg2005, Chevalier2006}. For non-relativistic SNR shocks,  $\epsilon_{\rm{e}} \approx K_{\rm{ep}}\xi_{\rm{cr}}$, where $K_{\rm{ep}}$ is the electron-to-proton ratio and $\xi_{\rm{cr}}$ is the acceleration efficiency of cosmic rays. Based on multi-wavelength spectra of Tycho, \protect\cite{Morlino2012} deduced $K_{\rm{ep}} = 1.6 \times 10^{-3}$ and $\xi_{\rm{cr}} \sim 0.06$, which implies $\epsilon_{\rm{e}} < 10^{-4}$. Similarly low values of $K_{\rm{ep}} \sim 10^{-3} - 10^{-4}$ have been deduced for young SNRs \protect\citep{Berezhko2006, Berezhko2009a, Berezhko2009},  while $\xi_{\rm{cr}} \lesssim 0.2$ for strong shocks characteristic of young SNRs \protect\citep{Caprioli2014}. Furthermore, particle-in-cell (PIC) simulations of electrons and ions in non-relativistic shocks recover $K_{\rm{ep}} \approx 10^{-3} - 10^{-2}$ for shock velocities $v_s/c \sim 0.01-0.1$ as seen in young SNRs \protect\citep{Park2015}. Recently, \protect\cite{BarniolDuran2016} deduced that for the \protect\cite{Badenes2010} LMC SNR sample, $\epsilon_{\rm{e}} \epsilon_{\rm{b}} \sim 10^{-3}$. This is higher than the values we obtain for the M33 sample, $\epsilon_{\rm{e}} \epsilon_{\rm{b}} \sim 10^{-3} - 10^{-5}$ (based on Figures \ref{fig:paramspace} and \ref{epsilonB}). This discrepancy could be a result of differences in model assumptions and completeness limits, and will be further explored in a future paper where we apply our model to the LMC SNRs recovered from deeper radio surveys (Huizenga et. al, in prep). 

Our result suggests that older SNRs are more efficient at accelerating electrons than younger ones. In fact, our estimate of $\epsilon_{\rm{e}}$ gives $K_{\rm{ep}} \sim \epsilon_{\rm{e}}/\xi_{\rm{cr}} \sim 0.04$, which is consistent with the value of $K_{\rm{ep}} \sim 0.01$ measured in the cosmic rays detected on Earth, which are thought to originate from SNRs with a wide variety of ages \protect\citep{Beringer2012, Morlino2012}. In addition, equipartition between electrons and magnetic field energies is neither assumed in our analysis, nor retrieved from it. A caveat in our measurement is that we do not consider the effect of orientation of the ambient magnetic field near the shock vicinity on the electron/proton acceleration \protect\citep{Reynoso2013, Caprioli2014, Caprioli2015}. The electron spectral index, $p$ can also shift the value of $\epsilon_{\rm{e}}$, but as discussed in Section \ref{caveat}, $\epsilon_{\rm{e}} \gtrsim 10^{-3}$ for reasonable variations in $p$, which implies a $K_{\rm{ep}} (\gtrsim 0.01)$ that is still comparable to cosmic-rays.  
\subsection{Missing SNRs and Superbubbles}
\label{sec:misssnr}
Our work implies that archival radio surveys of Local Group galaxies have missed a significant fraction of SNRs as shown in Figure \ref{fig:lumdiam}. For the best-fit parameters and our simple ISM model, nearly $30-40 \%$ of the simulated population of SNRs in M33 falls below the 3$\sigma$ detection limit of \protect\cite{Gordon1999} survey. Most of these SNRs are Type Ia, which explode farther from the mid-plane than CC SNRs and encounter lower densities on average. In fact SN1006, a well-known, young, low surface brightness Galactic Type Ia SNR that evolved in a low ISM density 550 pc from the Galactic mid-plane \protect\citep{Winkler2003, Berezhko2012} would have been missed by the survey at M33. Future radio surveys with deeper sensitivity limits will be able to increase the SNR inventory in the Local Group and provide us with a larger sample of objects for calculating the Local Group DTD.

We suspect that the largest radio-selected SNRs in M33 (ones with diameters > 80 pc in Figure \ref{fig:lumdiam}) may be superbubbles \protect\citep{MacLow1988} due to their unusually large luminosities and diameters. Even with the best-fit parameter values and observationally constrained distribution of SNR kinetic energies and ambient densities, SNRs rarely occur in this region of luminosity-diameter space. \protect\cite{Long2010} echoed the same concern that several SNRs in their X-ray catalogs may be superbubbles, as even our Galaxy, there are no known SNRs larger than 90 pc. It is beyond the scope of our current study to prove that the largest SNRs in Figure \ref{fig:lumdiam} are indeed, super-bubbles. \protect\cite{Gordon1999} warned that some of the largest SNRs in the catalog may be associated with or embedded in HII regions, causing over-estimated optical diameters. \protect\cite{Long2010} also concede that super-bubbles are mainly powered by ionizing radiation of OB stars, and therefore unlikely to create high [SII/H$\alpha$] ratios that would be indicative of shocked gas, as seen in SNRs. For this work, we do not expect superbubbles to affect the statistical results since there are so few of them compared to the total SNR sample of M33. In a future version of our model, we will include a simple treatment of super-bubbles and reinvestigate the observed SNR distribution.

\subsection{Effects of changing key assumptions} \label{caveat}
Our parameter constraints depend on certain assumptions made in our model. The electron spectral index, $p$ can vary between 2.2 and 2.4 \protect\citep{Caprioli2012} with steeper electron spectra yielding lower radio luminosities because fewer GeV electrons contribute to the synchrotron emission. In general, for diffusive shock acceleration (DSA), the electron spectra is expected to steepen with weakening shocks. We found a $\sim 25\%$ increase in log $\epsilon_{\rm{e}}$ when varying $p$ from 2.1 to 2.5, but without any significant change in the SN rate. The distribution of kinetic energies, which is uncertain for SNe, also affects the radio LFs. While Type Ia and CC both deposit a similar amount of energy ($\sim 10^{51}$ ergs) in our model, CC SNe are known to be intrinsically fainter than Type Ia in SN surveys \protect\citep{Li2011a}. Based on this, if we instead assume CC SNe deposit a mean energy of $10^{50.5}$ ergs, log $\epsilon_{\rm{e}}$ increases by 25$\%$ since the model tries to compensate for production of fainter SNRs. Increasing $\sigma_{\rm{log E}}$ for CC SNe from 0.1 to 0.4 causes the radio LFs to be brighter by $\sim 3\sigma$, where $\sigma$ is the shaded uncertainty regions of the LFs in Figure \ref{fig:compare}. Despite these variations, the value of $\epsilon_{\rm{e}}$ remains above $10^{-3}$, which is still 10 times higher than conventional estimates for young SNRs.

Our assumption of $z_*$ for Type Ia and CC SNe are based on observations in the Milky Way, but they may be different for M33. For the best fit values of $R$, $z_0$ and $\epsilon_{\rm{e}}$, we found that changing the scale height of Type Ia has little effect on our parameters since only 25$\%$ of the SNe produced in our model are Type Ia. But changing the CC scale height, e.g. from 90 to 200 pc, increases the SN rate by $\sim 30\%$, with a negligible increase in $\epsilon_{\rm{e}}$.

We therefore checked the effects of changing $R$, $z_0$ and $\epsilon_{\rm{e}}$ on the visibility time. Out of the three parameters, $z_0$ has the strongest effect on the visibility time since it controls the volume densities where SNRs explode for a given value of $N_H$. For example, changing $z_0$ from 200 to 400 pc changes the correlation, $t_{\rm{vis}} \propto N^{-a}_H$ in Figure. \ref{fig:vistime} from $a=0.33$ to $a=0.44$, and the fraction of detection-limited SNRs from 30$\%$ to $\approx 42\%$. In comparison, increasing $R$ by a factor of 6 had a negligible effect on $a$ or the fraction of detection-limited visibility times, but increasing $\epsilon_{\rm{e}}$ by a factor of 2 increased the luminosities of SNRs by a factor of 2, and reduced the fraction of detection-limited visibility times rom $30\%$ to $18\%$. In subsequent papers, we will explore ways to reduce the errors, both statistical and systematic, in our parameter constraints and incorporate them into our calculation of the Local Group DTD. 

\section{Conclusions}
We introduced a semi-analytical model capable of reproducing the radio luminosity function of SNRs in Local Group galaxies, taking into account the measured ISM distribution of the galaxy, the physics of SNR evolution and synchrotron emission, the diversity of SN explosions and the detection limits of a radio survey. Using this model, we can obtain observationally constrained estimates of radio visibility times of an SNR catalog in a galaxy - a critical ingredient in calculating SN rates and DTDs using SNRs. We applied our model to the M33 radio SNR catalog of \protect\cite{Gordon1999} and \protect\cite{Chomiuk2009a}, and derived the following essential results, 
\begin{enumerate}
\item M33 has an estimated SN rate $\sim 3.1 \times 10^{-3}$ SN per year, or roughly 1 SN every 320 years. The measurement is unique since it combines physical modelling of SNR shocks and constraints from an SNR catalog with a well-defined completeness limit \protect\citep{Chomiuk2009a}, and is therefore relatively robust to observational limitations. 
\item We measured the electron acceleration efficiency by SNR shocks, $\epsilon_{\rm{e}} \sim 4.2 \times 10^{-3}$ in an SNR sample dominated by SNRs in the Sedov stage, using the current models of field amplification in SNR shocks and taking into account the possibility of missing faint SNRs in the sample. Our estimate of the electron-to-proton ratio, $K_{\rm{ep}} \sim 0.04$ is consistent with measurements in cosmic rays detected on earth, and much higher than in young, ejecta-dominated SNRs.

\item The model predicts a correlation between the radio luminosity of the brightest SNR and the SN rate, similar to the prediction by \protect\cite{Chomiuk2009a} and \protect\cite{Chomiuk2011} on the correlation between the brightest SNR and the star-formation rate. The correlation roughly agrees with measurements of the brightest SNR and SN rates in nearby galaxies, and therefore serves as a consistency check for our model.

\item On average, about $30-40\%$ of our simulated SNRs in M33 fall below the detection limit of \protect\cite{Gordon1999}, given the assumptions in our model and our best-fit parameters. Most of the SNRs above the detection limit consists of core-collapse SNRs, whereas the missing SNRs are mostly Type Ia, which evolve in lower ambient densities and have lower surface-brightnesses. 

\item The radio visibility times ($t_{\rm{vis}}$) of $\sim 70\%$ SNRs in M33 are determined by their transition to the radiative phase, with characteristic timescales of 20-80 kyrs and a correlation with the ISM column density ($N_H$) in which they explode, i.e., $t_{\rm{vis}} \propto N_H^{-a}$, with $a \sim 0.33$. About $30\%$ of the SNRs will have shorter visibility times, determined by the detection limit of the radio survey. 
\end{enumerate} 

\section{Acknowledgements}
We are grateful to Anatoly Spitkovsky for his suggestions on the treatment of magnetic field amplification, Jeffrey Newman for his suggestions on the statistical methods, and Robert Braun for giving us permission to use the opacity-corrected HI column maps. We also thank Dan Maoz for his comments on the SNR visibility times and acknowledge helpful discussions with Christopher Kochanek and Rodolfo Barniol Duran. Lastly, we thank the anonymous referee for the constructive comments on our manuscript. SKS, LC, and CB acknowledge NSF/AST-1412980 for support of this work. This research has made use of NASA's Astrophysics Data System (ADS) Bibliographic Services, as well as the NASA/IPAC Extragalactic Data base (NED). The National Radio Astronomy Observatory is a facility of the National Science Foundation operated under cooperative agreement by Associated Universities, Inc., and the Westerbork Synthesis Radio Telescope is operated by the ASTRON (Netherlands Institute for Radio Astronomy) with support from the Netherlands Foundation for Scientific Research (NWO)

\label{lastpage}
\bibliographystyle{mnras}
\bibliography{Sarbadhicary2016_arXiv}

\begin{appendix}
\section{Radio Light Curve Model} \label{app:rlc}
\subsection{Energy spectrum of accelerated electrons}
Classically, acceleration of ions and electrons by SNRs has been described by the theory of diffusive shock acceleration (DSA). According to DSA, injected particles gain energy by being repeatedly scattered across the shock front by strong magnetic turbulence \protect\citep{ Axford1977, Krymskii1977, Bell1978, Blandford1978} The mechanism naturally predicts a power-law spectra for the accelerated electrons that resembles the cosmic ray spectrum observed on earth \protect\citep{Caprioli2015}. The theory was subsequently modified to account for the effects of streaming particles on the shock structure, the particle spectrum and the scattering of the particles upstream \protect\citep{Drury1983, Blandford1987, Jones1991, Malkov2001}. Evidence of electron DSA in SNRs has been gleaned from the featureless, synchrotron X-ray emission in the shells of young SNRs such as SN1006 \protect\citep{Koyama1995}, Tycho \protect\citep{Warren2005} and Cas A \protect\citep{Stage2006} as well as gamma ray emission in SNRs \protect\citep[e.g.][]{Morlino2012, Ackermann2013, Slane2014} which also provided evidence of efficient hadron acceleration.

Based on predictions of DSA, we assume the non-thermal synchrotron-emitting electrons are described by a power-law, 
\begin{equation} \label{eq:crlaw}
N(E) = N_0 E^{-p}
\end{equation}
where we choose $p = 2.2$ in our model, since the cosmic ray spectrum in some SNRs shows a deviation from the classical DSA spectrum with $p$ = 2 \protect\citep{Caprioli2012}. The normalization $N_0$ can be written as, 
\begin{equation} \label{N0}
N_0 = (p - 2)\ \epsilon_{\rm{e}}\ \rho_0\ v_s^2\ E_m^{(p - 2)}
\end{equation}
by assuming the average energy density of the electrons accelerated above a minimum energy $E_m\ ( = m_e c^2)$ is a constant fraction $\epsilon_{\rm{e}}$ of the post-shock energy density $\sim \rho_0 v_s^2$ \protect\citep{Soderberg2005, Chevalier2006}, where $\rho_0$ is the ISM density ($\rm{g/cm^{-3}}$) and $v_s$ is the shock velocity. Since $N_0 \propto v_s^2$, we see that the number of available synchrotron emitting electrons at any time during the SNR lifetime strongly varies with the shock velocity. 

\subsection{Magnetic Field Amplification} \label{app:mfa}
SNR shocks can amplify the surrounding magnetic field by over 2 orders of magnitude. This was evident in observations of thin X-ray rims around historical SNRs \protect\citep{Bamba2005, Warren2005, Vlk2005, Parizot2006}, which implied the presence of strong magnetic fields to confine synchrotron-emitting electrons near the shock vicinity. \emph{Chandra} observations of the pre-shock region of SN1006 suggests that this amplification must be induced in the upstream \protect\citep{Morlino2010}. The alternative scenario where damping of the magnetic field in the shock downstream produces the thin X-ray rims \protect\citep{Pohl2005} is inconsistent with the frequency dependence of the rim widths \protect\citep{Ressler2014}.
\begin{table*}
\centering
\begin{tabular}{|lll|}
\hline
\multicolumn{3}{|c|}{Ejecta-dominated} \\
\hline\hline
  & Shock Radius, $R_s(t) = $ & Shock velocity, $v_s(t) = $ \\
\hline
Type Ia & $(1.29\ \mathrm{pc})\ t_2^{0.7}\ E_{\mathrm{51}}^{0.35}\ n_0^{-0.1}\ M_{\mathrm{ej}}^{-0.25}$ & $(8797\ \mathrm{km/s})\ t_2^{-0.3}\ E_{\mathrm{51}}^{0.35}\ n_0^{-0.1}\ M_{\mathrm{ej}}^{-0.25}$ \\\\
Core Collapse & $(1.26\ \mathrm{pc})\ t_2^{0.75}\ E_{\mathrm{51}}^{0.38}\ n_0^{-0.08}\ M_{\mathrm{ej}}^{-0.29}$ & $(9213\ \mathrm{km/s})\ t_2^{-0.25}\ E_{\mathrm{51}}^{0.38}\ n_0^{-0.08}\ M_{\mathrm{ej}}^{-0.29}$  \\
\hline\hline
\multicolumn{3}{|c|}{Sedov Taylor}\\
\hline
\hline
Type Ia/ & $(12.5\ \mathrm{pc})\ t_4^{0.4}\ E_{\mathrm{51}}^{0.2}\ n_0^{-0.2}$ & $(490\ \mathrm{km/s})\ t_4^{-0.6}\ E_{\mathrm{51}}^{0.2}\ n_0^{-0.2}$ \\
Core Collapse & & \\
\hline
\end{tabular}
\caption{Expressions for shock radius and velocity for Type Ia and CC SNRs based on \protect\cite{Truelove1999}. We define $t_2 = t/(100\ \rm{yrs})$, $t_4 = t/(10^4\ \rm{yrs})$, $E_{\rm{51}} = E/(10^{51} \rm{ergs})$ as the kinetic energy of explosion, $M_{\rm{ej}} = M/ (1 \rm{\rm{M_{\odot}}})$ as the ejecta mass and $n_0$ as the ambient medium density in units of $\rm{1 cm^{-3}}$. The onset of Sedov-Taylor phase happens at $t = t^* (\mathrm{423\ years}) E_{51}^{-1/2} M_{\mathrm{ej}}^{5/6} n_0^{-1/3}$, where $t^* = 0.481$ for Type Ia, and $0.424$ for CC SNRs.}
\label{table:rs}
\end{table*}

Field amplification results from excitation of unstable wave modes by the streaming of accelerated particles ahead of the shock. These modes may be resonant with the Larmor radius of the accelerated particles \protect\citep{Bell1978} or non-resonant with shorter wavelengths that grow faster than resonant modes \protect\citep{Bell2004}. During earlier stages, i.e. free expansion and early Sedov-Taylor, when the shock velocity is high and particle acceleration is efficient, substantial field amplification is caused by non-resonant modes, whereas resonant Bell modes become important to the field amplification as the SNR becomes older and the shock weakens \protect\citep{Amato2009, Caprioli2014a}. We split the calculation of the amplified field into these two regimes.
\begin{figure} 
\includegraphics[width=\columnwidth]{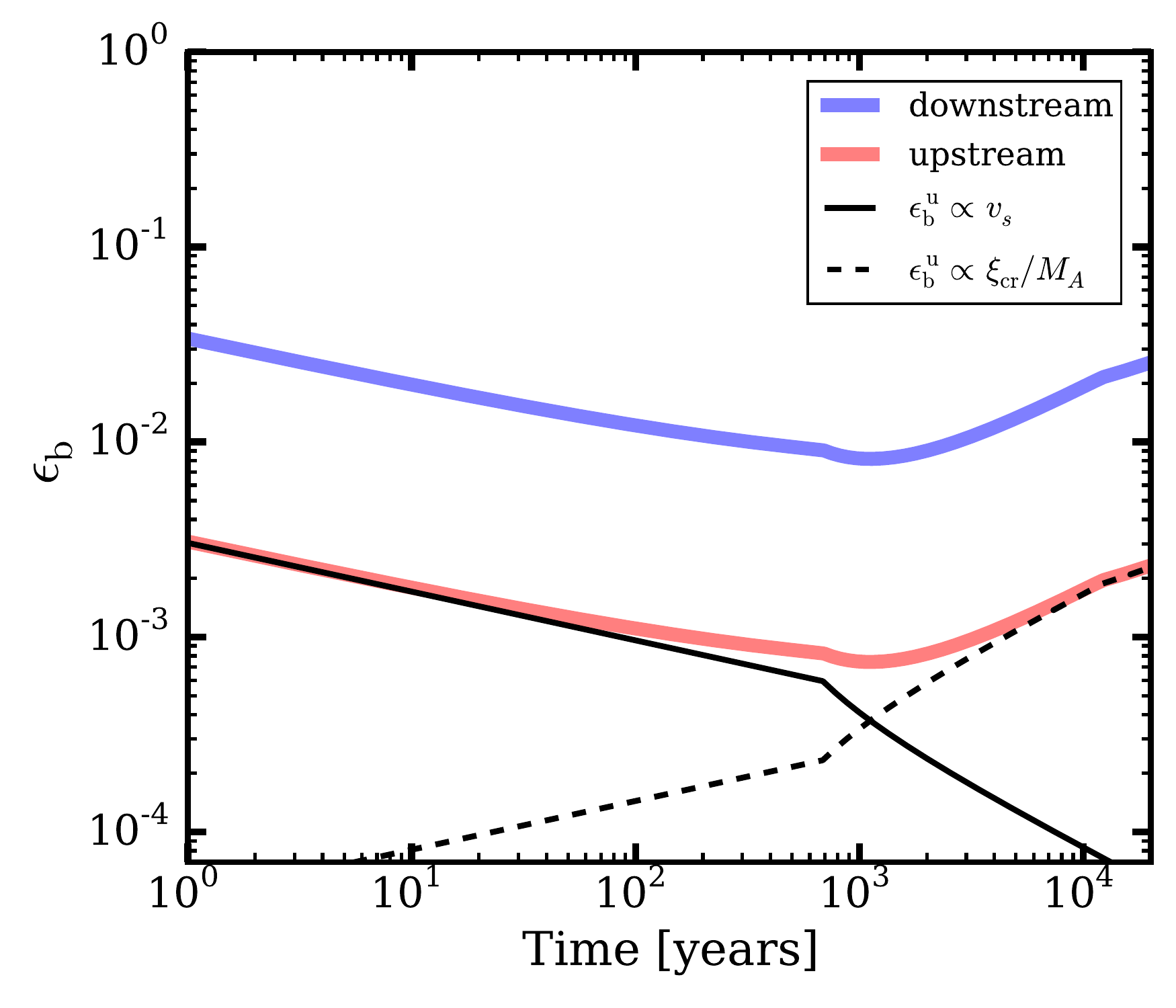}
\caption{Time evolution of $\epsilon_{\rm{b}}$ in the upstream (denoted as $\epsilon_{\rm{b}}^{\rm{u}}$ in the text) and in the downstream for an SNR with 5 $\rm{M_{\odot}}$ ejecta and $10^{51}$ ergs evolving into $n_0 = 1$ $\rm{cm^{-3}}$. The solid black line represent field amplification characterized by non-resonant wave modes, and dashed line represent amplification contributed by mostly resonant Bell modes. The red band shows the combined solution for $\epsilon_{\rm{b}}^{\rm{u}}$ we use in the model. The downstream value, shown with the blue band, is produced by compression of the upstream field and its values are consistent with estimates for known SNRs \protect\citep{Vlk2005}.The kink in the curves is the transition from the ejecta-dominated to the Sedov-Taylor solution (Table \ref{table:rs})}
\label{epsilonB}
\end{figure}
The energy density of the amplified upstream magnetic field, $B_u$ can be written as a fraction, $\epsilon^{\mathrm{u}}_{\mathrm{b}}$ of the shock energy density, 
\begin{equation} \label{eq:bu}
\epsilon^{\mathrm{u}}_{\mathrm{b}} = \frac{B^2_u/8 \pi}{\rho_0 v_s^2}
\end{equation}
We can scale $\epsilon^{\mathrm{u}}_{\mathrm{b}}$ with the Alfv\'{e}n Mach number of the shock, $M_A$ and the efficiency of particle acceleration, $\xi_{\mathrm{cr}}$, defined as the fraction of the shock energy in cosmic rays (ions + electrons). We define $M_A = v_s/v_A$, where $v_A$ is the Alfv\'{e}n velocity, given by $v_A = B_0/\sqrt{4 \pi \rho_0}$. Following \protect\cite{Crutcher1999}, the unshocked, background ISM magnetic field $B_0$ is given by, 
\begin{equation} \label{eq:B0}
B_0 = 9 \mathrm{\mu G} \left(\dfrac{\rho_0}{1.6 \times 10^{-27}\ \mathrm{g\ cm^{-3}}}\right)^{0.47}
\end{equation}
For high $M_A$ ($\gtrsim 100$) where non-resonant modes dominate, \protect\cite{Bell2004} argued that the amplified field saturates to a value, $B^2/8 \pi \sim 1/2 (v_s/c) \xi_{\mathrm{cr}}\rho_0 v_s^2$, due to increasing tension in the field lines. Using Eq \eqref{eq:bu}, we can write $\epsilon^{\mathrm{u}}_{\mathrm{b}}$ as,
\begin{equation} \label{eq:nonres}
\epsilon^{\mathrm{u}}_{\mathrm{b}} = \frac{1}{2} \left(\frac{v_s}{c}\right) \xi_{\mathrm{cr}}
\end{equation}
For field amplification with significant contribution from resonant modes, \protect\cite{Caprioli2014a} showed that the amplification scales as $B^2/B_0^2 \approx \xi_{\mathrm{cr}}M_A$. Using the definition of and Eq. \eqref{eq:bu}, we can write
\begin{equation} \label{eq:res}
\epsilon^{\mathrm{u}}_{\mathrm{b}} = \frac{1}{2} \frac{\xi_{\mathrm{cr}}}{M_A}
\end{equation}
Considering the dual roles of resonant and non-resonant streaming instabilities, we assume $\epsilon^{\mathrm{u}}_{\mathrm{b}}$ for our model SNRs of the form, 
\begin{equation} \label{eq:epsu}
\epsilon^{\mathrm{u}}_{\mathrm{b}} = \frac{\xi_{\mathrm{cr}}}{2} \left(\frac{v_s}{c} + \frac{1}{M_A}\right)
\end{equation}
We assume $\xi_{\mathrm{cr}} = 0.1$ for all SNRs, but the cosmic ray acceleration efficiency has been shown to decrease in weakened shocks \protect\citep{Caprioli2014a}. Based on this result, we set $\xi_{\mathrm{cr}}/10^{-2} = 0.15 M_A + 6$ for shocks with $M_A \lesssim 30$. 

The downstream magnetic field, $B$ is amplified by compression of the upstream field. Assuming $B_u$ is isotropic, and only the transverse components are compressed, $B$ can be written as,
\begin{equation} \label{eq:bdown}
B = \sqrt{\frac{1 + 2\eta^2}{3}} B_u
\end{equation}
where $\eta$ is the compression ratio $\approx 4 M^2/(M^2 + 2)$ given by the Rankine-Hugoniot jump conditions for a strong, non-radiative shock, where $M$ is the magneto-sonic Mach number. Since our SNR shocks have $v_s \geq 200$ km/s, they will have $M \gtrsim 10$. throughout. As a result, $\eta \approx 4$ throughout the SNR lifetime.

The evolution of $\epsilon^{\mathrm{u}}_{\mathrm{b}}$ is shown in Figure \ref{epsilonB}, and sets the general shape of our radio light curves in the Sedov-Taylor stage. The range of values spanned by the downstream $\epsilon_{\mathrm{b}} = \epsilon_{\mathrm{b}}^{\mathrm{u}} (1 + 2 \eta^2)/3$ is consistent with estimates in the downstream shock region of shell-type SNRs \protect\citep{Vlk2005}.

\subsection{SNR dynamics} \label{app:dyn}
We follow the work of \protect\cite{Truelove1999} which provides analytical expressions for the shock radius and velocity for the ejecta-dominated and Sedov-Taylor phases that closely match numerical results. In this formalism, each SNR starts as an explosion within a very small volume, depositing kinetic energy $E_{\rm{51}}$ (in units of $10^{51}$ ergs) and ejected mass $M_{\rm{ej}}$ which drives a spherically symmetric shock into a uniform ISM with number density $n_0$ (These quantities are selected for each SNR by the Monte-Carlo scheme described in Section \ref{MCSNR}). The resulting SNR consists of a forward shock, followed by a layer of shocked ISM, and a `contact discontinuity' that separates these outer layers from inner layers of shocked ejecta and the reverse shock. We are only interested in the forward shock since its larger volume and higher velocity generates most of the synchrotron emission \protect\citep{Chevalier2006}. 
 
The interaction of the power law SNR ejecta ($\rho \propto v_s^{-n}$) with a uniform ambient medium gives rise to $R_s \propto t^{(n-3)/n}$ and $v_s \propto t^{-3/n}$ in the ejecta-dominated phase that smoothly connects with the self-similar Sedov-Taylor solution, $R_s \sim t^{2/5}$ and $v_s \propto t^{-3/5}$ in the TM99 model. This feature allows us to generate an analytical light curve that spans the SNR lifetime up to the radiative phase, where we assume the synchrotron emission shuts off. We select $n=10$ for Type Ia SN, which describes a stellar envelope with polytrope = 4/3, such as a WD envelope, expelled by the SN shock   \protect\citep{Matzner1999, Chomiuk2016} and $n=12$ for core collapse (CC) SNRs \protect\citep{Chevalier1982a}. 

Table \ref{table:rs} lists the scaling relations for $R_s$ and $v_s$ in the ejecta-dominated and Sedov-Taylor phases. Both Type Ia and CC SNR ejecta expand rapidly during the ejecta-dominated phase. However, CC SNRs expand slightly faster than Type Ia's, because the steeper density profile of CC ejecta puts more energy per unit mass in the outer ejecta near the forward shock, allowing it to be decelerated at a slower pace by the ISM. At around 423 years, $R_s$ and $v_s$ varies according to the Sedov solution, which is the same for both Type Ia and CC SNRs since they have no memory of the initial ejecta mass at this stage. This causes the Sedov-Taylor light curves to be the same for both Type Ia and CC SNRs. 
\subsection{Radio Luminosity}
We follow the circumstellar interaction model of radio synchrotron emission of \protect\cite{Chevalier1998} where emission at lower frequencies is assumed to be suppressed by synchrotron self-absorption. Although some SN light curves are better described by free-free emission \protect\citep{Panagia2006}, the choice of optically thick emission process in our case is irrelevant because SNRs are already optically thin. We consider an SNR as the stage when the radio emission turns on as a result of circumstellar interaction of the ejecta, in contrast with intermediate-age radio supernovae \protect\citep{Cowan1985, Stockdale2001}. 

In the \protect\cite{Chevalier1998} model, the SNR radio emission region is approximated as a projected disc in the sky with radius $R$ and thickness $s$ with the same emitting volume, 
\begin{equation} \label{seq}
\pi R_s^2 s = \epsilon_f \left( \frac{4}{3} \pi R_s^3\right)
\end{equation}
where $\epsilon_f \approx 0.38$ is the emission filling factor \protect\citep{Chomiuk2016}. The synchrotron luminosity emitted from such a configuration is given as, 
\begin{equation} \label{lum:main}
L_{\nu} = 4 \pi^2 R_s^2 \left(\frac{c_5}{c_6}\right) B^{-1/2} \left[1 - \mathrm{exp}\left\{-\left(\frac{\nu}{\nu_1}\right)^{-(p+4)/2}\right\}\right] \left(\frac{\nu}{2 c_1}\right)^{5/2}
\end{equation}
where $R_s$ is the SNR radius, $B$ is the downstream magnetic field amplified by cosmic-ray induced instabilities, $N_0$ and $\gamma$ are parameters of the electron spectrum in Eq. \eqref{eq:crlaw}, $\nu$ is the frequency and $c_5 = 9.68 \times 10^{-21}$,  $c_6 = 8.1 \times 10^{-41}$ and $c_1 = 6.27 \times 10^{18}$ in cgs units are constants \protect\citep{Pacholczyk1970}. The frequency at which the SNR ejecta transitions from optically thick to thin is,
\begin{equation} \label{nu1}
\nu_1 = 2 c_1 (s c_6 N_0)^{2/(p+4)} B^{(p+2)/(p+4)}
\end{equation}
Assuming SNRs have optically thin ejecta ($\nu >> \nu_1$) at $\nu=1.4$ GHz, and using the aforementioned definitions of $B$, $N_0$ and $\nu_1$, we can rewrite the 1.4 GHz luminosity defined in Eq. \eqref{lum:main} in cgs units as,
\begin{equation} 
\begin{split}
L_{1.4}\ \approx\ & (2.2 \times 10^{24}\ \mathrm{ergs/s/Hz}) \\
& \left(\frac{R_{\rm{s}}}{10\ \mathrm{pc}}\right)^3 \left(\frac{\epsilon_{\rm{e}}}{10^{-2}}\right) \left(\frac{\epsilon^{\mathrm{u}}_{\mathrm{b}}}{10^{-2}}\right)^{0.8} \left(\frac{v_{\rm{s}}}{500\ \rm{km/s}}\right)^{3.6} \\
\end{split}
\end{equation}

\section{Model-Data Comparison using Maximum Likelihood} \label{app:maxlik}
We compare the radio LFs of the observed and model SNRs using a maximum likelihood method prescribed in \protect\cite{Badenes2010}, with similar implementations in \protect\cite{Maoz1993} and \protect\cite{Badenes2012a}. Each LF is divided into several bins, and we assume the probability of finding $j$ SNRs in the $i$th bin, for which the model predicts $n_i$ SNRs is a Poisson distribution,
\begin{equation}
P(j|n_i) = \dfrac{\mathrm{e}^{-n_i} n_i^j}{j!}
\end{equation}
Because binning the LFs can result in loss of information, we checked the effects of increasing the number of bins and found that the constraints on our parameter space in Section \ref{sec:paramspace} are almost unchanged even if the number of bins are doubled.

We assumed a Poisson likelihood since very few SNRs will occupy the bright end of the LF. We account for the fluctuation in the number of SNRs every year in the steady state by taking several snapshots of the population at different ages and obtaining an aggregate of LFs, which is compared with observations. Therefore, $n_i$ the mean number of SNRs per luminosity bin. In order to reduce statistical errors in $n_i$, we generate the steady state at a higher SN rate (e.g. $\sim 20\ R$), and then scale down the LF, and therefore the statistical errors in $n_i$, by the same factor. The likelihood of the model is thus given by,
\begin{equation}
\mathrm{ln}\ \mathcal{L}(R, z_0, \epsilon_{\rm{e}}) = \sum_{i=1}^{N_{bins}} \mathrm{ln}\ P(j|n_i)
\end{equation}

We used this likelihood function to compute the probability distributions of the respective parameters by marginalizing over the nuisance parameters, as shown by the shaded regions and histograms in Figure \ref{fig:paramspace}. We sampled 10 data points per parameter to recover the marginal probabilities and then interpolated between these values to construct the probability contours. This was done because of the sizable computational time required for generating each steady state SNR population for a given vector of parameters. Because of this, the peaks of our marginalized probabilities may slightly fluctuate, but the areas of parameter space ruled out by our model remain stable. 
\end{appendix}
\end{document}